\documentclass[aps,prc,reprint,superscriptaddress]{revtex4-1}

\usepackage{graphicx}
\usepackage{dcolumn}
\usepackage{amsmath} 
\usepackage{physics}
\usepackage{subfloat}
\usepackage{bm}

\graphicspath{{plots/}}

\usepackage{tikz}
\usetikzlibrary{patterns,arrows,calc,decorations.pathmorphing}
\tikzset{
    level/.style      = {ultra thick, black},
    wlevel/.style    = {thin, black},
    virtual/.style    = {thick, densely dashed},
    trans/.style      = {thick,->,shorten >=1pt,shorten <=1pt,>=stealth,sloped,below}, 
    wtrans/.style    = {->,shorten >=1pt,shorten <=1pt,>=stealth,sloped,below},    
    strans/.style    = {very thick,->,shorten >=1pt,shorten <=1pt,>=stealth,sloped,below}, 
    sstrans/.style    = {ultra thick,->,shorten >=1pt,shorten <=1pt,>=stealth,sloped,below}, 
    vtrans/.style    = {dashed,->,shorten >=1pt,shorten <=1pt,>=stealth,sloped,below},    
    mtrans/.style    = {thick,->,shorten >=1pt,shorten <=1pt,>=stealth,sloped,decorate,decoration={snake,amplitude=1.5}},    
    connect/.style = {dashed,black},
    notice/.style    = {draw,rectangle callout,callout relative pointer={#1}},
    label/.style      = {text width=2cm},
    specfac_s1/.style      = {ultra thick, red},
    specfac_d3/.style      = {ultra thick, black!30!green},
    specfac_d5/.style      = {ultra thick, cyan},    
    specfac_g7/.style      = {ultra thick, blue},
    specfac_g7_3/.style  = {ultra thick, magenta},
    nucleon/.style={circle, draw, fill=black},    
    neutron/.style={circle, draw, fill=red},    
    proton/.style={circle, draw, fill=blue},        
    mint/.style    = {blue,thick,sloped,decorate,decoration={snake,amplitude=1.5}} 
}

\newcommand{\nuc}[2]{$^{#1}$#2}

\newcommand{\zps}{$0^{+}$ states}
\newcommand{\dpsr}[1]{$d(^{#1}$Sr$,p)$}
\newcommand{\dpsrf}[2]{$d(^{#1}$Sr$,p)^{#2}$Sr}

\newcommand{\ppsr}[1]{$p(^{#1}$Sr$,p)$}
\newcommand{\ddsr}[1]{$d(^{#1}$Sr$,d)$}
\newcommand{\sfac}{spectroscopic factor}
\newcommand{\sfacs}{spectroscopic factors}

\begin{document}



\title{Single-Particle Structure of Neutron-Rich Sr Isotopes Via \boldmath{$d(^{94,95,96}$Sr$,p)$} Reactions}

\author{S.~Cruz}
\affiliation{Department of Physics and Astronomy, University of British Columbia, Vancouver, BC V6T 1Z4, Canada}
\affiliation{TRIUMF, Vancouver, BC V6T 2A3, Canada}
\author{K.~Wimmer}
\email{Corresponding author: wimmer@phys.s.u-tokyo.ac.jp}
\affiliation{Department of Physics, Central Michigan University, Mt Pleasant, MI 48859, USA}
\affiliation{Department of Physics, The University of Tokyo, Hongo, Bunkyo-ku, Tokyo 113-0033, Japan}
\author{P.C.~Bender}
\affiliation{TRIUMF, Vancouver, BC V6T 2A3, Canada}
\author{R.~Kr\"{u}cken}
\affiliation{Department of Physics and Astronomy, University of British Columbia, Vancouver, BC V6T 1Z4, Canada}
\affiliation{TRIUMF, Vancouver, BC V6T 2A3, Canada}
\author{G.~Hackman}
\affiliation{TRIUMF, Vancouver, BC V6T 2A3, Canada}
\author{F.~Ames}
\affiliation{TRIUMF, Vancouver, BC V6T 2A3, Canada}
\author{C.~Andreoiu}
\affiliation{Department of Chemistry, Simon Fraser University, Burnaby, BC V5A 1S6, Canada}
\author{R.A.E.~Austin}
\affiliation{Department of Astronomy and Physics, Saint Mary's University, Halifax, NS B3H 3C2, Canada}
\author{C.S.~Bancroft}
\affiliation{Department of Physics, Central Michigan University, Mt Pleasant, MI 48859, USA}
\author{R.~Braid}
\affiliation{Department of Physics, Colorado School of Mines, Golden, CO 80401, USA}
\author{T.~Bruhn}
\affiliation{TRIUMF, Vancouver, BC V6T 2A3, Canada}
\author{W.N.~Catford}
\affiliation{Department of Physics, University of Surrey, Guildford, Surrey, GU2 7XH, United Kingdom}
\author{A.~Cheeseman}
\affiliation{TRIUMF, Vancouver, BC V6T 2A3, Canada}
\author{A.~Chester}
\affiliation{Department of Physics, Simon Fraser University, Burnaby, BC V5A 1S6, Canada}
\author{D.S.~Cross}
\affiliation{Department of Chemistry, Simon Fraser University, Burnaby, BC V5A 1S6, Canada}
\author{C.Aa.~Diget}
\affiliation{Department of Physics, University of York, York, YO10 5DD, United Kingdom}
\author{T.~Drake}
\affiliation{Department of Physics, University of Toronto, Toronto, ON M5S 1A7, Canada}
\author{A.B.~Garnsworthy}
\affiliation{TRIUMF, Vancouver, BC V6T 2A3, Canada}
\author{R.~Kanungo}
\affiliation{TRIUMF, Vancouver, BC V6T 2A3, Canada}
\affiliation{Department of Astronomy and Physics, Saint Mary's University, Halifax, NS B3H 3C2, Canada}
\author{A.~Knapton}
\affiliation{Department of Physics, University of Surrey, Guildford, Surrey, GU2 7XH, United Kingdom}
\author{W.~Korten}
\affiliation{IRFU, CEA, Universit\'{e} Paris-Saclay, F-91191 Gif-sur-Yvette, France}
\affiliation{TRIUMF, Vancouver, BC V6T 2A3, Canada}
\author{K.~Kuhn}
\affiliation{Department of Physics, Colorado School of Mines, Golden, CO 80401, USA}
\author{J.~Lassen}
\author{R.~Laxdal}
\author{M.~Marchetto}
\affiliation{TRIUMF, Vancouver, BC V6T 2A3, Canada}
\author{A.~Matta}
\affiliation{Department of Physics, University of Surrey, Guildford, Surrey, GU2 7XH, United Kingdom}
\affiliation{LPC, ENSICAEN, CNRS/IN2P3, UNICAEN, Normandie Universit\'{e}, 14050 Caen cedex, France}
\author{D.~Miller}
\author{M.~Moukaddam}
\affiliation{TRIUMF, Vancouver, BC V6T 2A3, Canada}
\author{N.A.~Orr}
\affiliation{LPC, ENSICAEN, CNRS/IN2P3, UNICAEN, Normandie Universit\'{e}, 14050 Caen cedex, France}
\author{N.~Sachmpazidi}
\affiliation{Department of Physics, Central Michigan University, Mt Pleasant, MI 48859, USA}
\author{A.~Sanetullaev}
\affiliation{Department of Astronomy and Physics, Saint Mary's University, Halifax, NS B3H 3C2, Canada}
\affiliation{TRIUMF, Vancouver, BC V6T 2A3, Canada}
\author{C.E.~Svensson}
\affiliation{Department of Physics, University of Guelph, Guelph, ON, N1G 2W1, Canada}
\author{N.~Terpstra}
\affiliation{Department of Physics, Central Michigan University, Mt Pleasant, MI 48859, USA}
\author{C.~Unsworth}
\affiliation{TRIUMF, Vancouver, BC V6T 2A3, Canada}
\author{P.J.~Voss}
\affiliation{Department of Chemistry, Simon Fraser University, Burnaby, BC V5A 1S6, Canada}

\date{\today}

\begin{abstract}
\begin{description}
\item[Background] The region around neutron number $N=60$ in the neutron-rich Sr and Zr nuclei is one of the most dramatic examples of a ground state shape transition from (near) spherical below $N=60$ to strongly deformed shapes in the heavier isotopes. 
\item[Purpose] The single-particle structure of \nuc{95-97}{Sr} approaching the ground state shape transition at \nuc{98}{Sr} has been investigated via single-neutron transfer reactions using the $(d,p)$ reaction in inverse kinematics. These reactions selectively populate states with a large overlap of the projectile ground state coupled to a neutron in a single-particle orbital.
\item[Method] Radioactive \nuc{94,95,96}{Sr} nuclei with energies of 5.5~$A$MeV were used to bombard a CD$_2$ target. Recoiling light charged particles and $\gamma$ rays were detected using a quasi-$4\pi$ silicon strip detector array and a 12 element Ge array. The excitation energy of states populated was reconstructed employing the missing mass method combined with $\gamma$-ray tagging and differential cross sections for final states were extracted. 
\item[Results] A reaction model analysis of the angular distributions allowed for firm spin assignments to be made for the low-lying 352, 556 and 681~keV excited states in \nuc{95}{Sr} and a constraint has been placed on the spin of the higher-lying 1666~keV state. Angular distributions have been extracted for 10 states populated in the \dpsrf{95}{96} reaction, and constraints have been provided for the spins and parities of several final states. Additionally, the 0, 167 and 522~keV states in \nuc{97}{Sr} were populated through the \dpsr{96} reaction. Spectroscopic factors for all three reactions were extracted.
\item[Conclusions] Results are compared to shell model calculations in several model spaces and the structure of low-lying states in \nuc{94}{Sr} and \nuc{95}{Sr} is well-described. The spectroscopic strength of the $0^+$ and $2^+$ states in \nuc{96}{Sr} is significantly more fragmented than predicted. The spectroscopic factors for the \dpsrf{96}{97} reaction suggest that the two lowest lying excited states have significant overlap with the weakly deformed ground state of \nuc{96}{Sr}, but the ground state of \nuc{97}{Sr} has a different structure.
\end{description}
\end{abstract}

\pacs{}
\keywords{single-particle structure, transfer reaction, shape coexistence}
\maketitle

\section{Introduction}
An atomic nucleus can deform its shape in order to minimize its energy. This is observed across the nuclear landscape, both in ground states and excited states. Indeed, it seems that even a small number of valence protons and neutrons outside of a closed core can drive the whole nucleus into a deformed shape. The long-range attractive residual proton-neutron ($p-n$) interaction allows the nucleus to gain additional binding energy by arranging the nucleons in certain ways across the valence orbitals, which in turn causes a departure from sphericity~\cite{Casten2000}. The expense of such re-arrangements is dependent on the size of the energy gaps between single-particle orbitals above the Fermi energy. 
If the energy spacing is small, the valence nucleons can scatter into valence orbitals which are above the Fermi energy and drive the nucleus into a low-energy deformed configuration. On the other hand, if the energy spacing is large, the valence nucleons are unable to scatter into higher orbitals and this favors spherical shapes. The size of these energy gaps is in turn dependent on the number of valence nucleons, due to the monopole component of the residual interaction. Clearly, the underlying shell structure of nuclei plays an important role in the propensity for nuclei to deform. 

The evolution of ground state shapes across an isotopic chain is commonly observed to be a gradual process, although in some cases the shape can change dramatically with the addition of just a few nucleons.
A striking example of this has been observed across the \nuc{}{Sr} and \nuc{}{Zr} isotopic chains, where an abrupt change of shape in the ground states takes place at $N\sim60$. The ground state shape transition has been measured directly using laser spectroscopy, as a sudden increase in charge radii at $N=60$~\cite{Buchinger1990}.
This is also evidenced by the sudden drop in $2^+_1$ energies across the even-even isotopes at $N\geq60$, which indicates that the ground state shape changes from a nearly spherical structure to a strongly deformed prolate ($\beta\sim0.4$) structure~\cite{ENSDF}. 
Recent Coulomb excitation measurements have established that the ground state of \nuc{96}{Sr} and the $0_2^+$ state in \nuc{98}{Sr} possess similar structures which, assuming axial symmetry, correspond to weakly deformed shapes with $\beta\sim0.1$~\cite{Clement2016,Clement2016}.  In the $N=56$ isotope \nuc{94}{Sr}, recent re-determination of the $B(E2; 2_1^+ \rightarrow 0_1^+)$ value from a lifetime measurement~\cite{Chester2017} supports the interpretation that the ground state in \nuc{94}{Sr} is close to spherical. Taken together, these measurements point towards a gradual evolution in shape up to $N\sim58$ with $\beta\leq0.1$ which then rapidly changes at $N=60$ to $\beta\sim0.4$ for the ground state. However, the degree of deformation in the ground state of the $N=59$ nucleus \nuc{97}{Sr} is not well understood although the spin and parity of the ground state has been established as $1/2^+$, which is not expected within the spherical shell model. The magnetic moments of the \nuc{95,97}{Sr} ground states were reported to be very similar through laser spectroscopy~\cite{Buchinger1990} and deviate from the shell model expectation.

 Also of interest is the emergence of shape-coexisting states in the vicinity of $N\sim60$ and $Z\sim40$. A very strong $E0$ transition between the 1229 and 1465~keV excited \zps\ in \nuc{96}{Sr}, with $\rho^2(E0)=0.185(50)$~\cite{JungThesis} is a strong indicator of mixing between states which have different intrinsic deformations. Enhanced $E0$ transition strengths between low-lying \zps\ have also been observed in the nearby nuclei \nuc{98}{Sr}, \nuc{98}{Zr}, \nuc{100}{Zr}, \nuc{100}{Mo} and \nuc{102}{Mo}~\cite{Wood1999}.
 
 The $N\sim60$, $Z\sim40$ region of the nuclear chart has been the subject of substantial interest theoretically for many years~\cite{Arseniev1969,FP1977,FP1979,Kumar1985,Michiaki1990,Skalski1993,Baran1995,Lalazissis1995,Skalski1997,Holt2000,Zhang2006,Rza209,RodGuz2010a,RodGuz2010,Sieja2009, Liu2011,Mei2012,Xiang2012,Petrovici2012,Togashi2016}.
It has been shown that the emergence of deformed low-energy configurations can be explained in the shell model by the evolution of single-particle structure and the interaction between protons and neutrons in certain valence orbitals, namely the spin-orbit partner orbitals $\pi0g_{9/2}$ and $\nu0g_{7/2}$~\cite{FP1977,FP1979}. State-of-the-art beyond mean field calculations have been able to reproduce the observed shape transition at $N=60$ in \nuc{}{Sr}, \nuc{}{Zr} and \nuc{}{Mo}~\cite{RodGuz2010, RodGuz2010a}, although correctly predicting the ground state spins and parities of the odd-mass isotopes remains a challenge. Ultimately, advances in theoretical models are limited by the experimental data that is available.
While numerous experiments have provided useful information on the \nuc{}{Sr} isotopes~\cite{Jung1980,Kratz1982,Buchinger1990,Mach1991,Wu2004,Park2016,Clement2016,Clement2016a,Regis2017}, 
a firm understanding of the underlying single-particle configurations of low-energy states is essential for a detailed description of this region
This situation motivated a series of single-neutron transfer reactions across the neutron-rich Sr isotopes \nuc{94,95,96}{Sr}. The main results for the \dpsr{95} reaction were already presented in~\cite{cruz18}. The present paper discusses the details of the experiment and the analysis as well as further results.

\section{Experimental setup and conditions}
The experiments were performed at the TRIUMF-ISAC-II facility~\cite{TRIUMF}. The \dpsr{94}\ and \dpsr{95,96}\ measurements were the first high mass (A$>$30) experiments with a re-accelerated secondary beam to be performed at TRIUMF. 
The Sr beams were produced by impinging a 480~MeV proton beam on a thick Uranium Carbide (UC$_x$) target. 
Sr atoms diffusing out of the UC$_x$ target were selectively ionized into a singly charged ($1^+$) state using the TRIUMF Resonant Ionization Laser Ion Source~\cite{TRIUMF} in order to enhance the extraction rate of the Sr species compared to surface-ionized contaminants, also produced within the production target.
The cocktail beam was then sent through the ISAC mass separator~\cite{TRIUMF} to produce a beam containing only isotopes of the same $A$ (94, 95, 96). The beam was then transported to the Charge State Booster where the isotopes were charge-bred by an Electron Cyclotron Resonance plasma source to a higher charge state (see Table~\ref{tab:BeamSummary} for details). This was necessary so that the beam could next be sent to the Radio-Frequency Quadrupole (RFQ), which accepts a maximum mass-to-charge ratio ($A/q$) of 30~\cite{TRIUMF}. Inside the RFQ, time-dependent electric fields were tuned to accelerate the specific $A/q$ of Sr ions. Contaminant isotopes in the beam were mismatched with the acceleration phase of the RFQ and so did not undergo any acceleration. Following the RFQ, these contaminants were deflected out of the beam using the bending dipole magnets in the accelerator chain. The beams were transported to the ISAC-II facility where their kinetic energy was increased to 5.5~$A$MeV using the superconducting linear accelerator~\cite{TRIUMF}.
Finally, the beams were transported to the experimental station where they impinged upon 0.5~mg/cm$^2$ deuterated polyethylene (CD$_2$) targets, mounted in the center of the SHARC silicon detector array~\cite{SHARC}. 

SHARC (Silicon Highly-segmented Array for Reactions and Coulex) is a compact arrangement of double-sided silicon strip detectors which is optimized for high geometrical efficiency and excellent spatial resolution, with $\Delta \theta_\text{lab} \sim 1 ^\circ$ and $\phi$ coverage of approximately 90\%. 
The SHARC array configuration consists of two double-sided silicon strip detector (DSSSD) box sections (DBOX and UBOX) and
an annular DSSSD detector (UQQQ). 
The downstream DBOX section, with the approximate angular range $35^\circ<\theta_\text{lab}<80^\circ$, was configured using a $\Delta E-E$ detector arrangement (140~$\mu m$ DSSSDs and 1~mm thick unsegmented pad detectors) so that different ions could be identified (Fig.~\ref{fig:KinematicsPid}). For scattering angles $\theta_\text{lab}<90^\circ$ elastic scattering of protons and deuterons overlaps with the kinematic lines of the transfer reactions requiring the particle identification.
In the upstream UBOX ($95^\circ<\theta_\text{lab}<140^\circ$) and UQQQ ($147^\circ<\theta_\text{lab}<172^\circ$) sections, particle identification was not used as only protons are emitted with $\theta_\text{lab}>90^\circ$ (as shown in Fig.~\ref{fig:KinematicsPid}). Background events arise from $\beta$ decay of radioactive beam accidentally stopped in the scattering chamber, and light particles emitted in fusion evaporation reactions with carbon in the CD$_2$ target. The former can be suppressed by the particle identification cut as shown in the inset of Fig.~\ref{fig:KinematicsPid} in laboratory forward direction and a cut on the detected energy in backward direction. Protons from fusion evaporation reactions contribute a continuous background to the excitation energy spectra. This background is more pronounced at laboratory forward angles due to the forward focusing of the reaction products. If unambiguous identification of the state populated in the reaction by $\gamma$-ray coincidences is possible the residual background is negligible.

The SHARC array was mounted in the center of the TIGRESS $\gamma$-ray detector array~\cite{TIGRESS}. In these experiments, TIGRESS was composed of 12 HPGe clover detectors arranged in a compact hemispherical arrangement with approximately $2\pi$ steradians geometrical coverage (see Fig. 2 of~\cite{matta19}). 
The individual crystals contain an electrical core contact and eight-fold electrical segmentation on the outer contact; four quadrants and a lateral divide, giving an overall 32-fold segmentation within each clover. This segmentation enhances the sensitivity to the emission angle of the $\gamma$ ray to enable more precise Doppler reconstruction. For transitions from states with very short lifetimes the in-beam resolution after Doppler corrections amounts to $0.6$~\%. The segmented design also made it possible to improve the quality of the data taken in TIGRESS by using add-back to reconstruct full $\gamma$-ray energies from multiple scattering events. The Compton suppressor shields were not used in the present work.

The beam composition was measured at regular intervals during the experiment using a Bragg ionization detector~\cite{TBRAGG}, which was positioned on another beam-line adjacent to the TIGRESS experimental station.
The beam composition in each experiment was also analyzed using $\beta$-decay data from the radioactive beam-like ions which were scattered onto the DQQQ (not instrumented in the present work). 
The primary contaminant in each beam were the isobars \nuc{94-96}{Rb}.
Contributions from non-isobaric $A/q$ contaminants, originating from the ISAC CSB, were found to be negligible in the $A=94$ and 95 beams. However, substantial \nuc{17}{O} contamination was identified in the first half of the $A=96$ beam-time due to challenges in beam tuning. Only the data taken during the second half of the $A=96$ beam time was analyzed.
Further details regarding the beam are given in Table~\ref{tab:BeamSummary}. 
\begin{table}[h]         
\begin{center}       
\def\arraystretch{1.3} \setlength\tabcolsep{5pt}
  \begin{tabular}{ccccc}
    \hline
    \hline    
   Beam  &  Q [e] & Rate [s$^{-1}$]$^*$  &  Duration [days]  &  Purity [\%]  \\
    \hline
    \hline    
     \nuc{94}{Sr} & $15^+$ & $\sim$ 3x10$^4$ & $\sim$ 3 & 50(5)\\
     \nuc{95}{Sr} & $16^+$ & $\sim$ 1.5x10$^6$ & $\sim$ 2.5 & 95(3)\\     
     \nuc{96}{Sr} & $17^+$ & $\sim$ 1x10$^4$ & $\sim$ 1 & 58(13)\\          
    \hline
    \hline        
    \multicolumn{5}{l}{$^*$\footnotesize{including contaminations}}\\
  \end{tabular}
  \caption{Summary of the \nuc{94,95,96}{Sr} beam properties. }
\label{tab:BeamSummary}  
\end{center}
\end{table}

\section{Analysis and results} 
\begin{figure}[h!]
\includegraphics[width=\columnwidth]{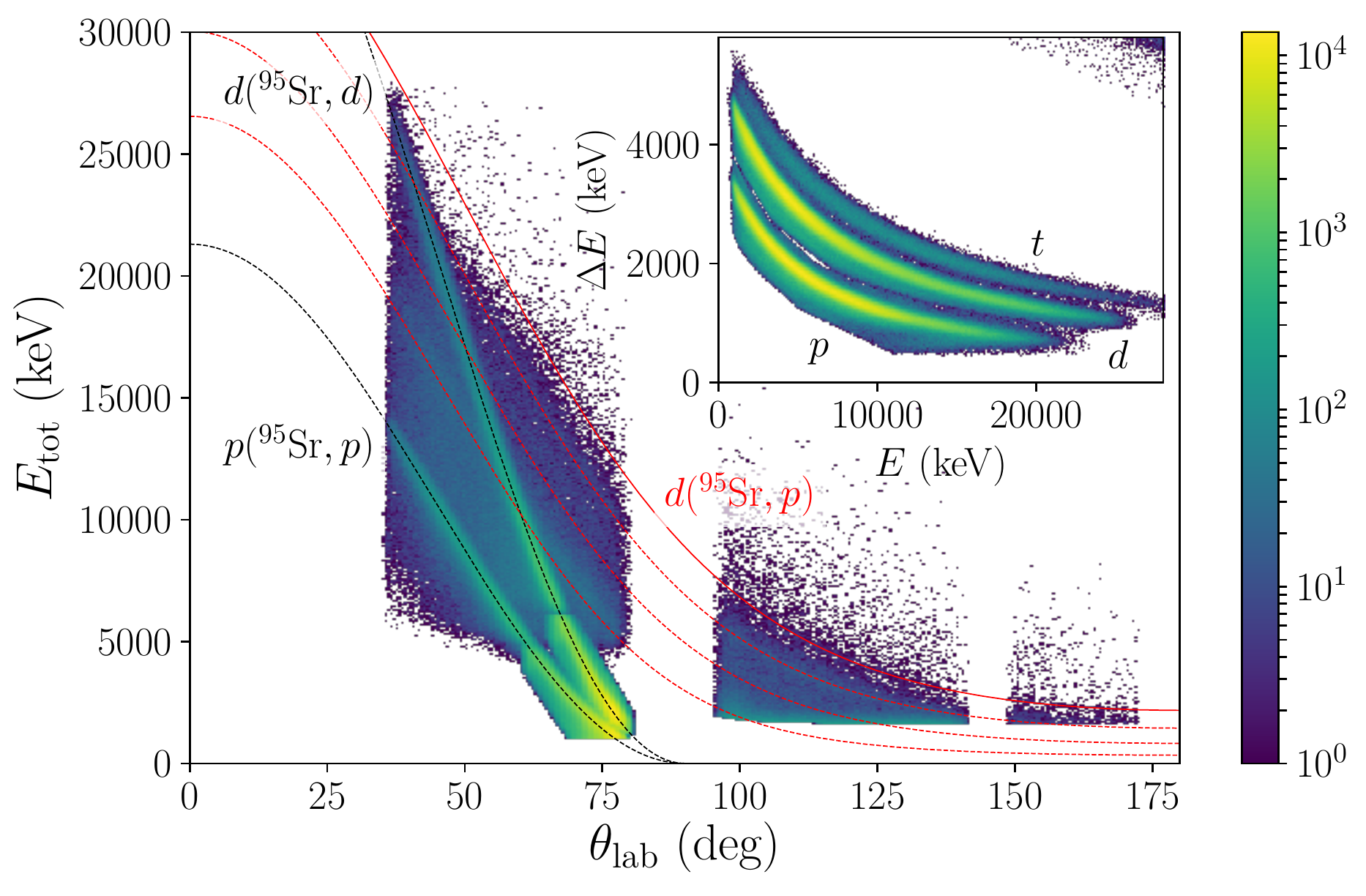}%
\caption{Kinematics plot for \nuc{95}{Sr} incident on the CD$_2$ target, compared to calculated kinematics lines drawn for elastic scattering (black, dotted lines) and (d,p) transfer at 0, 2, 4 and 6~MeV excitation energy (red). In addition to uniquely identified particle in the DBOX, elastic scattered protons and deuterons are shown below the identification threshold of about 5000~keV identified by their kinematic $E(\theta_\text{lab})$ relation. The inset shows the particle identification plot for the DBOX section (see text), which was used to distinguish between protons and deuterons. }
\label{fig:KinematicsPid}
\end{figure}

The SHARC and TIGRESS detectors were calibrated using standard sources. In the case of TIGRESS \nuc{60}{Co} and \nuc{152}{Eu} sources were used to obtain the energy and efficiency calibrations of each detector. The $\Delta E$ detectors of SHARC were calibrated using a triple alpha source. The $E$ detectors were calibrated using the proton and deuteron elastic scattering data, which was acquired simultaneously with the \dpsr{} data. Fig.~\ref{fig:KinematicsPid} shows the kinetic energy of measured protons and deuterons as a function of laboratory scattering angle for the \nuc{95}{Sr} beam incident on the CD$_2$ target. 
The total kinetic energy of measured particles was reconstructed by adding calculated energy losses using SRIM~\cite{Srim2010} in the target and Si detector dead layers to the energy deposited in SHARC. The energy loss correction amounted less than 100 keV for protons in laboratory forward direction as well as for scattering angles larger than 120$^\circ$, and up to 500 keV for protons scattered close to 100$^\circ$.
Details of the calibration methods can be found in ref.~\cite{CruzThesis}. 
The excitation energy ($E_\text{x}$) was reconstructed using the measured energy and scattering angle of the detected particles using the missing mass method.
 The excitation energy resolution of the DBOX, UBOX and UQQQ sections was determined to be approximately 550, 450 and 400~keV (FWHM) for the respective angular ranges. The primary contributions to the energy resolution were the energy loss of the beam and proton recoils in the thick target. For this reason, excited states which were less than approximately 500~keV apart could not be individually resolved. Excited states were thus identified using the de-excitation $\gamma$ ray in addition to an $E_\text{x}$ gate~\cite{catford10}. For low statistics cases, such as the \nuc{94}{Sr} and \nuc{96}{Sr} experiments, a constrained multi-peak fit was used to consistently extract the population strengths of unresolved adjacent states. This is discussed further in the subsequent sections.

\begin{figure}[h!]
\includegraphics[width=\columnwidth]{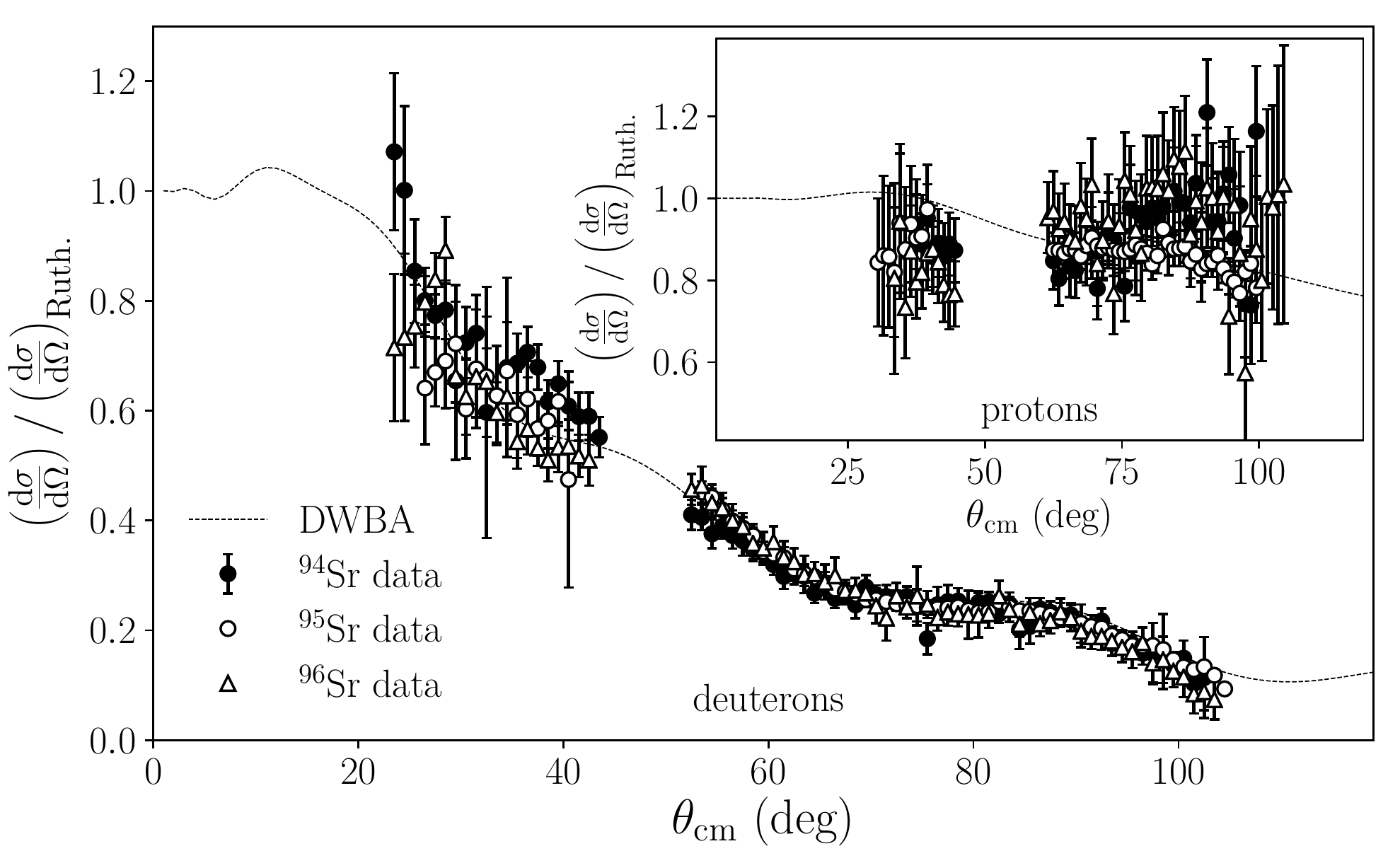}%
\caption{Comparison of \ddsr{94,95,96} angular distribution data to DWBA calculations using the optimized optical potential that is given in Table~\ref{tab:OpticalModel}. The inset shows the comparison of the \ppsr{94,95,96} data to the global potential PP-76~\cite{Perey1976} (see text).}
\label{fig:Elastics}
\end{figure}

 The experimental angular distributions were compared to distorted wave Born approximation (DWBA) calculations that were carried out using the FRESCO code~\cite{FRESCO}. The optical model parameters used in the analysis were determined from fits to the elastic scattering data measured simultaneously. For the proton optical potential the data are not sensitive to the parameters and the parametrization of ref.~\cite{Perey1976} was used in the following. Several global optical model parameter sets~\cite{Perey1976,Daehnick1980,LHOpticalModel} were compared to the $(d,d)$ angular distributions and it was found that the parameters of Lohr and Haeberli~\cite{LHOpticalModel}, with some small adjustments, resulted in very good agreement with the combined $(d,d)$ data for all three experiments. The combined fit for \ddsr{94,95,96} can be seen in Fig.~\ref{fig:Elastics}.
It should be noted that the angular distributions shown in Fig.~\ref{fig:Elastics} include the contributions for the beam contamination (mainly Rb), however the parameters are expected to vary slowly with $A$ and $Z$. The parameters used in the analysis of the transfer reaction data are summarized in Table~\ref{tab:OpticalModel}.
 \begin{table*}[t!]                 
\begin{center}
\def\arraystretch{1.3} \setlength\tabcolsep{10pt}
  \begin{tabular}{c|cccccccccc}
    \hline
    \hline
    Data & $R_{c}$ & $V_{0}$ & $R_{0}$ & $A_{0}$& $W_{D}$ & $R_{D}$ & $A_{D}$ &  $V_{SO}$ & $R_{SO}$ & $A_{SO}$ \\
    \hline
    \hline    
    (d,d), This Work & 1.30 & 109.45 & 1.07 & 0.86 & 10.42 & 1.37 & 0.88 & 7.00 & 0.75 & 0.50 \\     
   (d,d), LH-74~\cite{LHOpticalModel} & 1.30 & 109.45 & 1.05 & 0.86 & 10.42 & 1.43 & 0.77 & 7.00 & 0.75 & 0.50 \\            
    \hline
    (p,p), PP-76~\cite{Perey1976} & 1.25 & 58.73 & 1.25 & 0.65 & 13.50 & 1.25 & 0.47 & 7.50 & 1.25 & 0.47 \\     
    \hline
    \hline
  \end{tabular}
  \caption{Optical model parameters that were used to describe \nuc{94,95,96}{Sr} elastic scattering angular distributions in the DWBA calculations (Fig.~\ref{fig:Elastics}). The global optical model parameters of Lohr and Haeberli (LH-74)~\cite{LHOpticalModel}, with some small adjustments were found to give the best fit to the combined $(d,d)$ data. The global optical model parameters of Perey and Perey (PP-76) were used to describe the combined $(p,p)$ data.}
  \label{tab:OpticalModel}
\end{center}
\end{table*}
The overall normalization constant, required to convert the experimental cross sections into units of mb/sr, was also determined from the elastic scattering. The ratio of proton and deuteron elastic scattering in each experiment was used to determine the fraction of deuterons and protons within the CD$_2$ target, 96(2)\%, 92(1)\% and 96(2)\% deuterons for the \nuc{94,95,96}{Sr} experiments, respectively. The uncertainties include statistical and reaction model uncertainties. The normalization constants were corrected for the beam purity and target deuteron content.

The \dpsr{94,95,96} reactions were modeled as a single-step process where the transferred neutron populates an unoccupied valence orbital. By comparing the experimental cross section for each final state to the calculations, the spectroscopic factor can be extracted. In addition to the statistical uncertainty, these spectroscopic factors carry a theoretical systematic uncertainty arising from the choice of the reaction model, optical model parameters, and the potential used to calculate the nucleon bound-state wave function. By comparing different parametrizations, this uncertainty has been estimated to be $20$~\%. Relative spectroscopic factors are not affected by the uncertainty. 
In order to better gauge the uncertainty arising from the reaction modeling, adiabatic distorted wave approximation (ADWA) calculations were also performed. For the incoming channel global nucleon-nucleus optical model parameters from~\cite{koning03} evaluated at half the beam energy were used. The ADWA model takes the breakup of the loosely bound deuteron explicitly into account, but the reliability at the rather low beam energies of the present work is not well established. In general the ADWA results describe the shape of the angular distribution better as shown below, and result in smaller spectroscopic factors by about 15\% compared to the DWBA.

By comparing the experimental angular distributions to reaction model calculations the most probable $\Delta\ell$ value was determined for each state using a $\chi^2$ analysis. It was not possible to differentiate between the spin-orbit partner orbitals $1d_{5/2}$ and $1d_{3/2}$ (both $\Delta \ell=2$), 
 and so both are given as possible scenarios where applicable.
 The neutron $0h_{11/2}$ ($\ell=5$) orbital was not considered here as the single-particle energy has been estimated as 3.5~MeV at \nuc{91}{Zr}~\cite{Sieja2009,Holt2000}.

\subsection{Results for the \dpsrf{94}{95} reaction}
\begin{figure}[h!]
\includegraphics[width=\columnwidth]{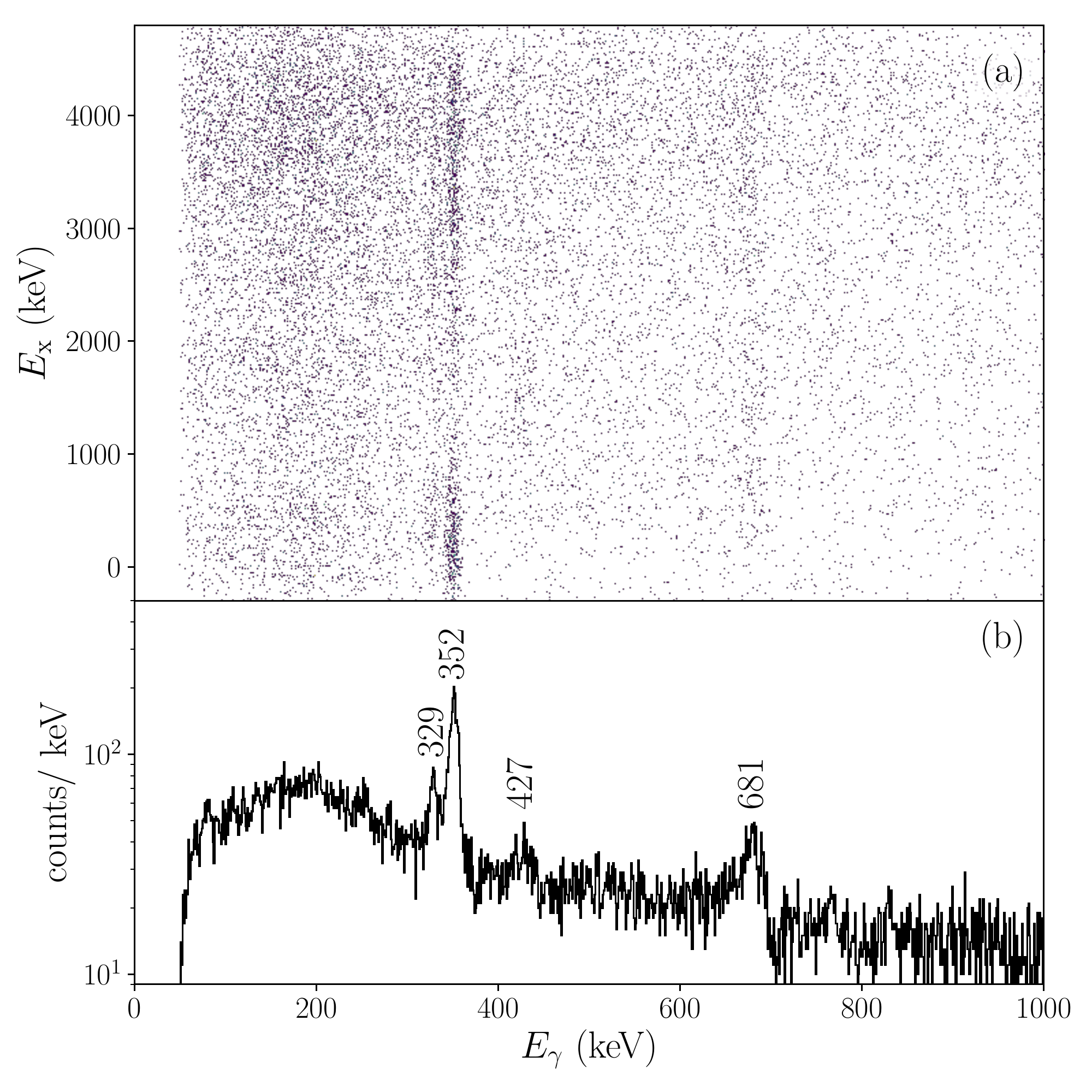}%
\caption{Excitation energy versus $\gamma$-ray energy matrix (upper) and projected $\gamma$-ray spectrum (lower panel) for \nuc{95}{Sr} states populated via \dpsr{94}.}
\label{fig:ExcGam95}
\end{figure}

\begin{figure}[h!]
  \includegraphics[height=0.85\columnwidth,angle=-90]{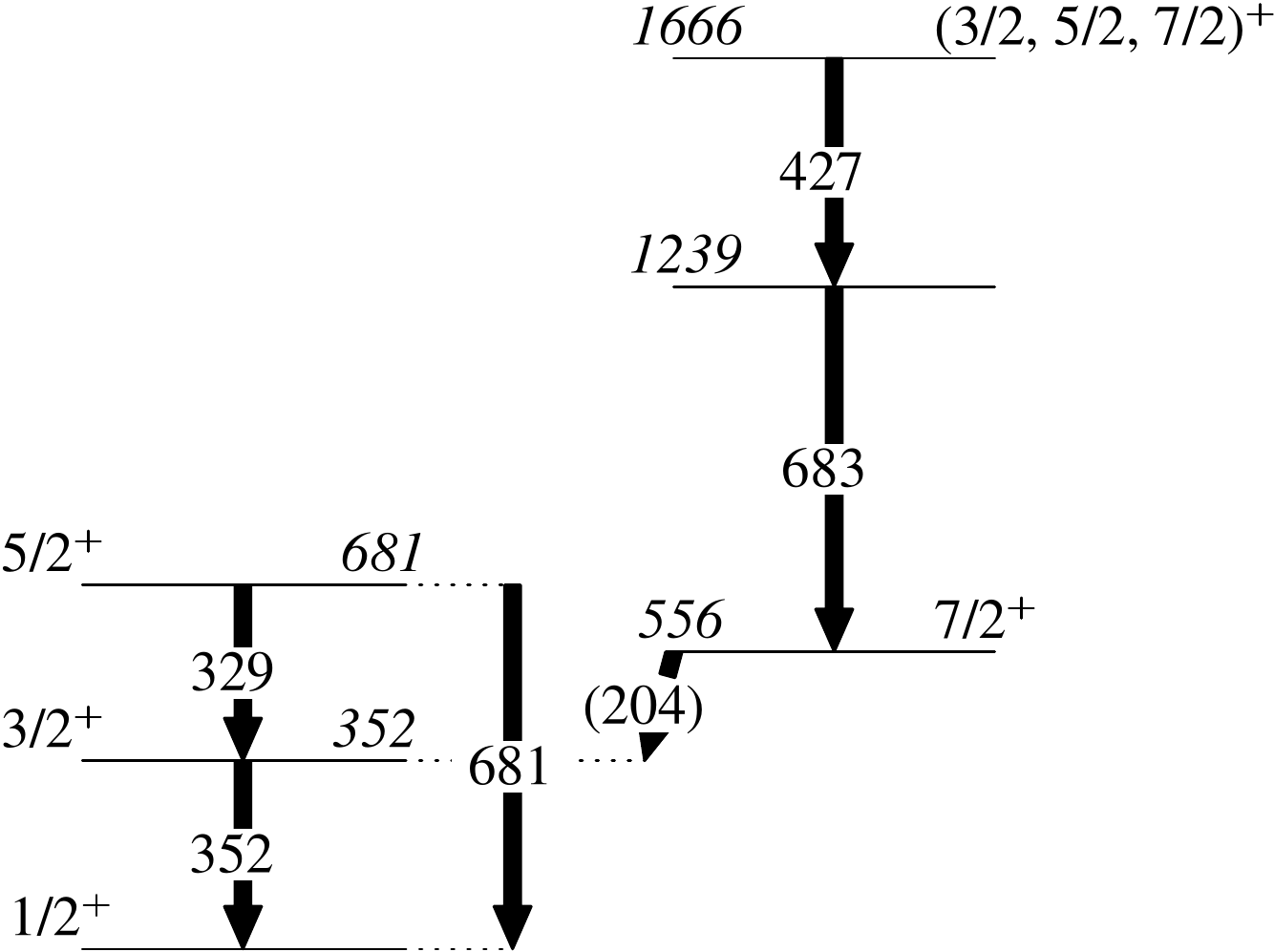}%
\caption{Level scheme for \nuc{95}{Sr} states that were populated through \dpsr{94}. The 204~keV $\gamma$ ray was not observed due to the 21.9(5) ns~\cite{Kratz1983,ENSDF} half-life of the 556~keV state (more details in the text).}
\label{fig:Levels95}
\end{figure}

The $\gamma$ rays and excitation energy of states in \nuc{95}{Sr} that were populated via the \dpsr{94} reaction are shown in Fig.~\ref{fig:ExcGam95}.
Strong 329, 352 and 681~keV $\gamma$-ray lines can be seen in the $E_\text{x}$ versus $E_\gamma$ matrix. Fig.~\ref{fig:Levels95} shows the \nuc{95}{Sr} level scheme for states that were identified below 2~MeV.
All states and transition energies were previously known. Substantial direct population of the 0, 352 and 681~keV states was observed. There is also clear evidence for the direct population of the 1666~keV excited state through the observation of the 427~keV $\gamma$ ray. This line is enhanced in the spectrum if a gate on excitation energies $1<E_\text{x}<2$~MeV is placed. However, the statistics were too low for an angular distribution analysis. It is also apparent that excited states up to $\sim$5~MeV were populated through this reaction and decay via the 352 and 681~keV states. However, it was not possible to identify any states above the 1666~keV state due to the limited statistics.

\begin{figure}[h]
\includegraphics[width=\columnwidth]{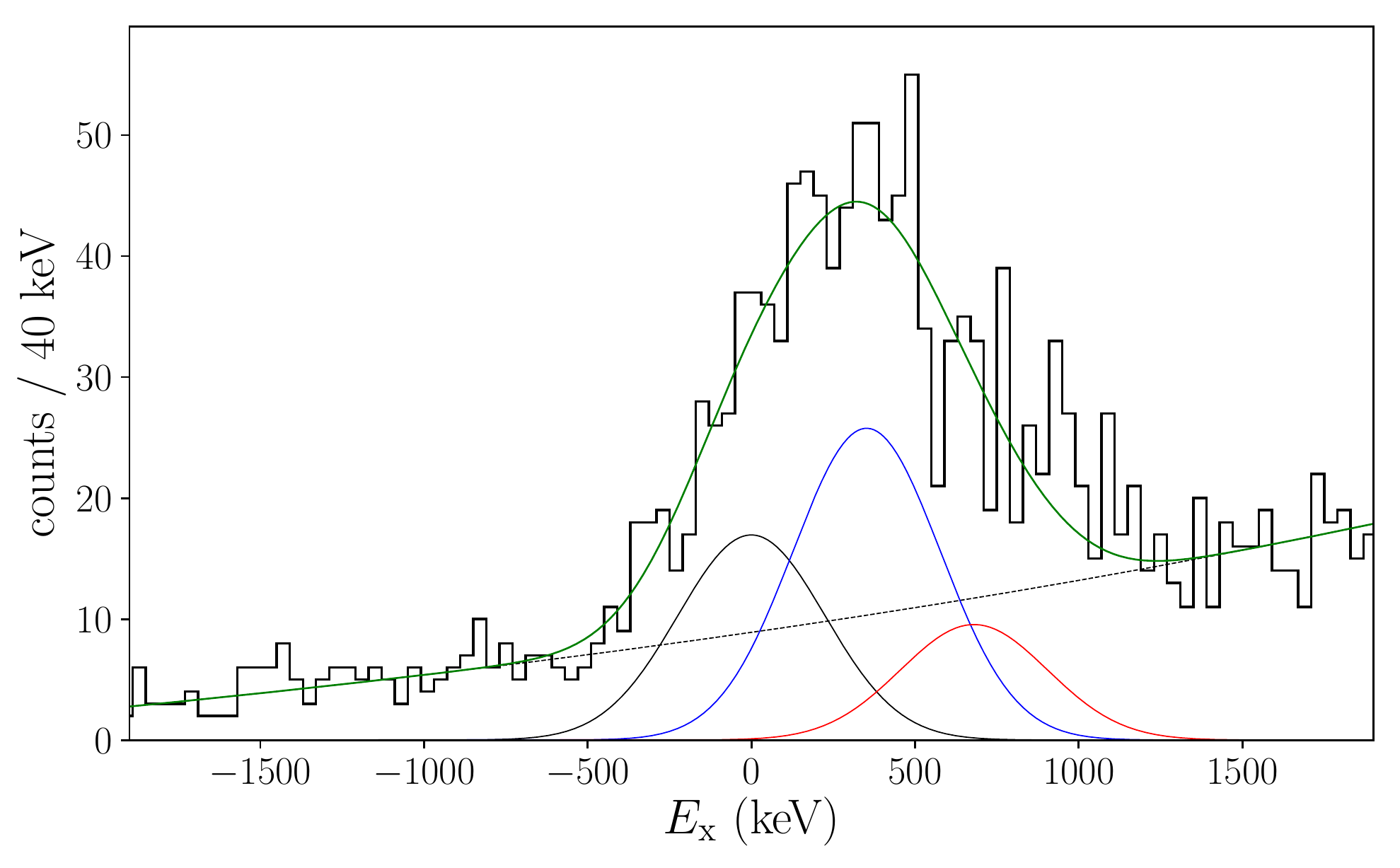}%
\caption{Excitation energy spectrum extracted from the recoiling proton energies and angles at a center of mass angle $\theta_\text{cm} = 30^\circ$. The continuous green line shows the constrained 3-peak fit of the 0, 352 and 681~keV \nuc{95}{Sr} states. The dashed line represents the continuous background.}
\label{fig:3PeakFit}
\end{figure}

\begin{figure}[h!]
  \includegraphics[width=\columnwidth]{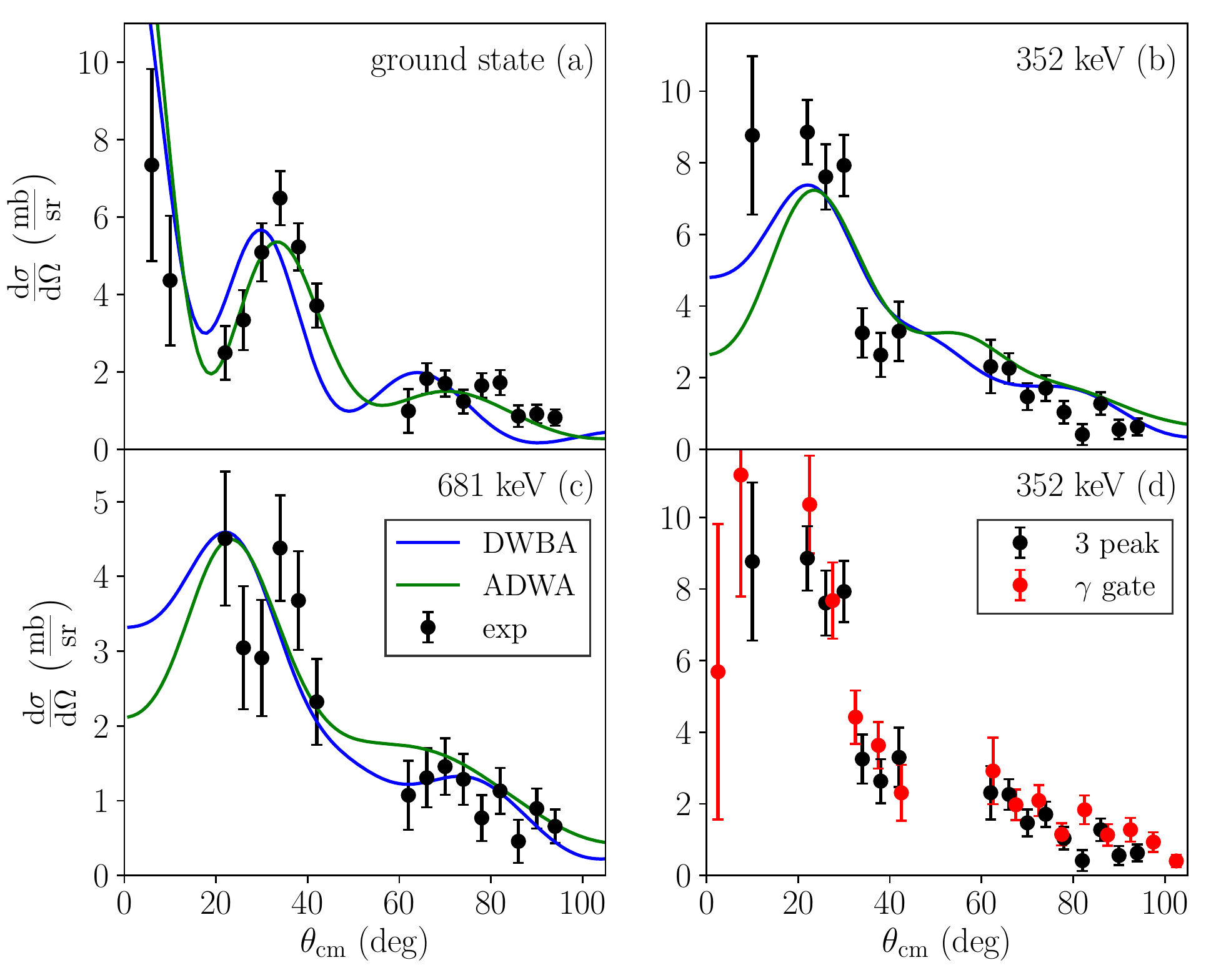}%
\caption{Panels (a-c): Comparison of the reaction model calculations to the angular distributions for the 0, 352 and 681~keV states in \nuc{95}{Sr}. The experimental data has been obtained from the constrained 3-peak fit (Fig.~\ref{fig:3PeakFit}). The solid lines are the best-fitting reaction model calculations using the DWBA (blue) and ADWA (green) methods. Panel (d): comparison of the two methods to extract the angular distribution for the 352~keV state (see text).}
\label{fig:AngDist95}
\end{figure}

\textit{The ground state of \nuc{95}{Sr}:}
The ground, 352, and 681~keV states were not clearly resolved in the excitation energy spectrum (Fig.~\ref{fig:3PeakFit}). Therefore the angular distributions were extracted simultaneously using a constrained three (Gaussian) peak-plus-exponential background fit of the excitation energy spectrum for each angular bin. An example fit is shown in Fig.~\ref{fig:3PeakFit}.
The peak widths and separations between them were fixed using the known $E_\text{x}$ resolution (determined with simulations and verified using the the \dpsr{95} data set~\cite{cruz18}) and the energies of the states, respectively. The shape of the ground state angular distribution (Fig.~\ref{fig:AngDist95} (a)) is in good agreement with the $\Delta\ell=0$ reaction model calculations, with a spectroscopic factor of 0.41(9) for the DWBA and 0.34(7) for the ADWA, respectively. Systematic uncertainties include the experimental sources discussed above and theoretical uncertainties arising from the optical model parameters used. 
Our results are thus consistent with the known $J^\pi = 1/2^+$ assignment for this state~\cite{Buchinger1987}. 

\textit{The 352~keV state:}
%
Two independent experimental angular distributions were produced for the 352~keV state; one was extracted using the three peak fit (see Fig.~\ref{fig:3PeakFit} (b)) and a second was extracted by gating on the 352~keV $\gamma$-ray transition and the excitation energy (Fig.~\ref{fig:AngDist95} (d)). The shape of both angular distributions are in clear agreement with the $\Delta\ell=2$ calculation, constraining the spin and parity of this state to be $J^\pi = 3/2^+$ or $5/2^+$. 
Combining the $\Delta \ell=2$ angular distribution with the previously established $M1$ character of the 352~keV $\gamma$-ray transition to the \nuc{95}{Sr} ground state~\cite{ENSDF} allows a firm spin and parity assignment of $3/2^+$ for this state. The spectroscopic factors for adding a neutron to the $1d_{3/2}$ orbital are 0.50(10) and 0.55(13), using the two methods respectively, using the DWBA reaction theory. The weighted average of the two spectroscopic factors is presented in Table~\ref{tab:Summary}. As for the ground state the ADWA calculation results in a slightly lower spectroscopic factor of 0.45(7).   

\textit{The 556~keV state:}
Although direct population of the long-lived 556~keV state ($T_\text{1/2} = 21.9(5)$~ns) in this experiment could not be confirmed owing to the low $\gamma$-ray detection efficiency due to its long lifetime, its spin and parity can be constrained by combining the $3/2^+$ assignment for the 352~keV state from this work with previous measurements.  
The 204~keV $\gamma$-ray transition from the 556~keV to the 352~keV state was previously determined to have pure $E2$ character using conversion electron spectroscopy~\cite{ENSDF}. Additionally no decay directly to the ground state has been observed in this or previous~\cite{ENSDF} work.
This constrains the spin and parity of the 556~keV state to be $J^\pi = 7/2^+$. 
The \dpsr{94} transfer reaction is not expected to populate $7/2^+$ states strongly as the large angular momentum transfer $\Delta\ell = 4$ suppresses the cross section. While no cross section or angular distribution could be extracted from the present data set, the spectrum in Fig.~\ref{fig:3PeakFit} shows that the direct population of this state must be small.

\textit{The 681~keV state:}
Three independent experimental angular distributions were produced for the 681~keV state. In addition to the three peak fit result (shown in Fig.~\ref{fig:AngDist95}), angular distributions (not shown) were also produced for this state by gating on the 329~keV and 681~keV transitions as well as the excitation energy. The shape of all three extracted angular distributions are in good agreement with each other and with the $\Delta\ell=2$ DWBA calculation, constraining the spin and parity of this state to be $J^\pi = 3/2^+$ or $5/2^+$. 
The absence of any $M1$ component in the 681~keV ground state transition~\cite{ENSDF} allows us to assign $J^\pi = 5/2^+$ to the 681~keV state.
The spectroscopic factors for population of the $1d_{5/2}$ orbital that were extracted (with the DWBA calculations) using the three methods are 0.20(5), 0.14(5) and 0.14(7), respectively. The weighted average of these spectroscopic factors is presented in Table~\ref{tab:Summary}. The ADWA analysis resulted in a weighted average spectroscopic factor of 0.14(3).

\textit{The 1666~keV state:}
The observation of a 427~keV peak in Fig.~\ref{fig:ExcGam95}, coincident with excitation energies in the range of $1<E_\text{x}<2$~MeV, establishes that the 1666~keV state was populated in the \dpsr{94} reaction. This state was observed in \nuc{252}{Cf} spontaneous fission decay~\cite{Hwang2004}, a process which preferentially populates high spin states. In that work a tentative spin and parity of $11/2^+$ was assigned based on the large branching ratio to the 1239~keV (tentative $9/2^+$) state. However, the population of the state in transfer makes this assignment unlikely. The addition of a single neutron to the \nuc{94}{Sr} ground state via the \dpsr{94} reaction can directly populate \nuc{95}{Sr} states with spins and parities of $1/2^+,3/2^+,5/2^+$, and $7/2^+$.  The cross section for $11/2^-$ states with $\Delta\ell=5$ is very low and is not further considered in this work. We therefore propose a spin and parity of $(3/2,5/2,7/2)^+$ for the 1666~keV state. The angular distribution for this state could not be extracted, comparison of the integrated cross section with the DWBA and ADWA calculations suggests a spectroscopic factor of $C^2S<0.05$ for $\Delta \ell =0, 2$ or $C^2S \approx 0.12$ for $\Delta \ell =4$ transfer to the $0g_{7/2}$ orbital.

\subsection{Results for the \dpsr{95} reaction}
\begin{figure}[h!]
\includegraphics[width=\columnwidth]{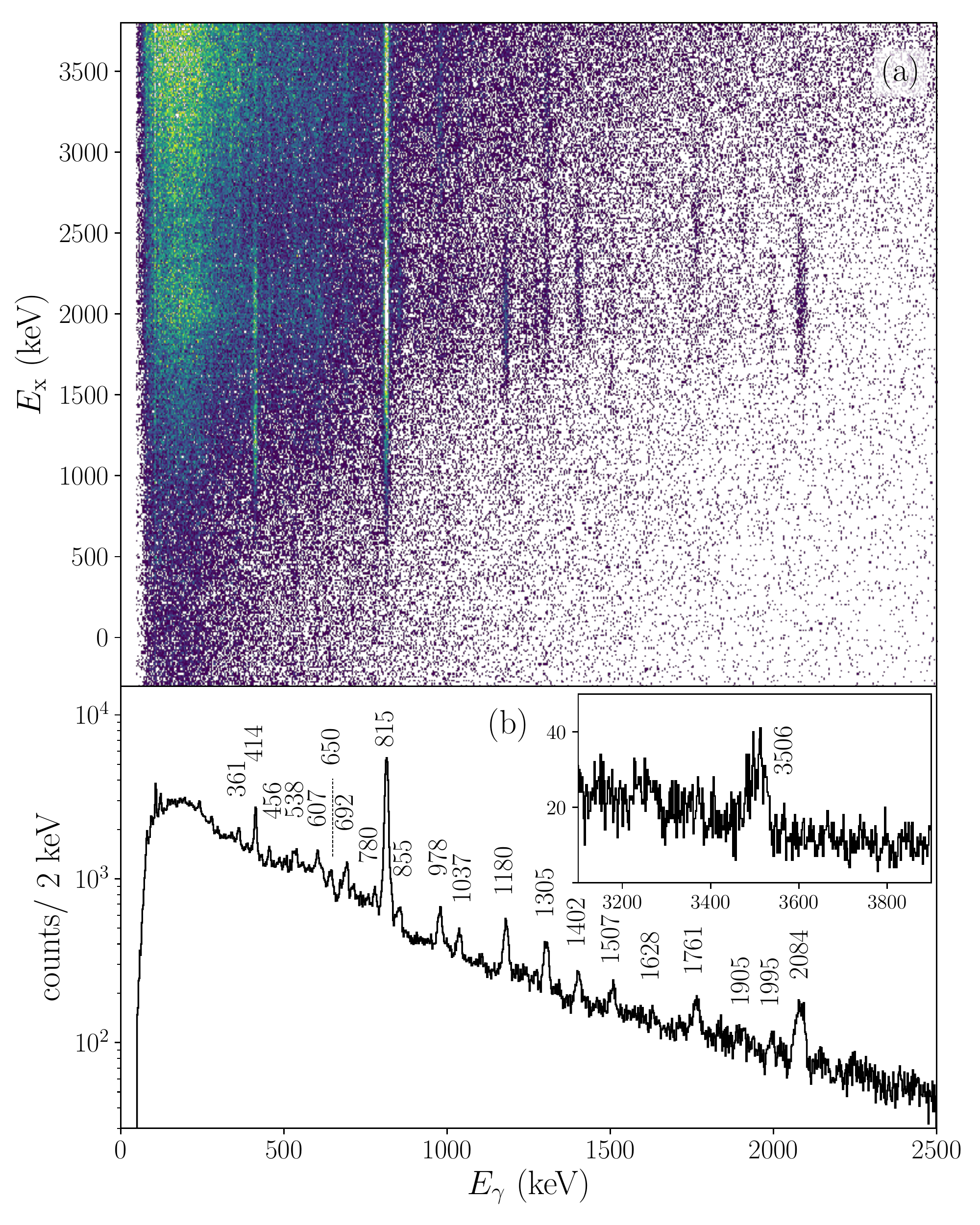}%
\caption{Excitation energy versus $\gamma$-ray energy matrix (upper) and projected $\gamma$-ray spectrum (lower) for \nuc{96}{Sr} states populated via the \dpsr{95} reaction.}
\label{fig:ExcGam96}
\end{figure}

\begin{figure*}[t!]
  \includegraphics[width=0.8\textwidth]{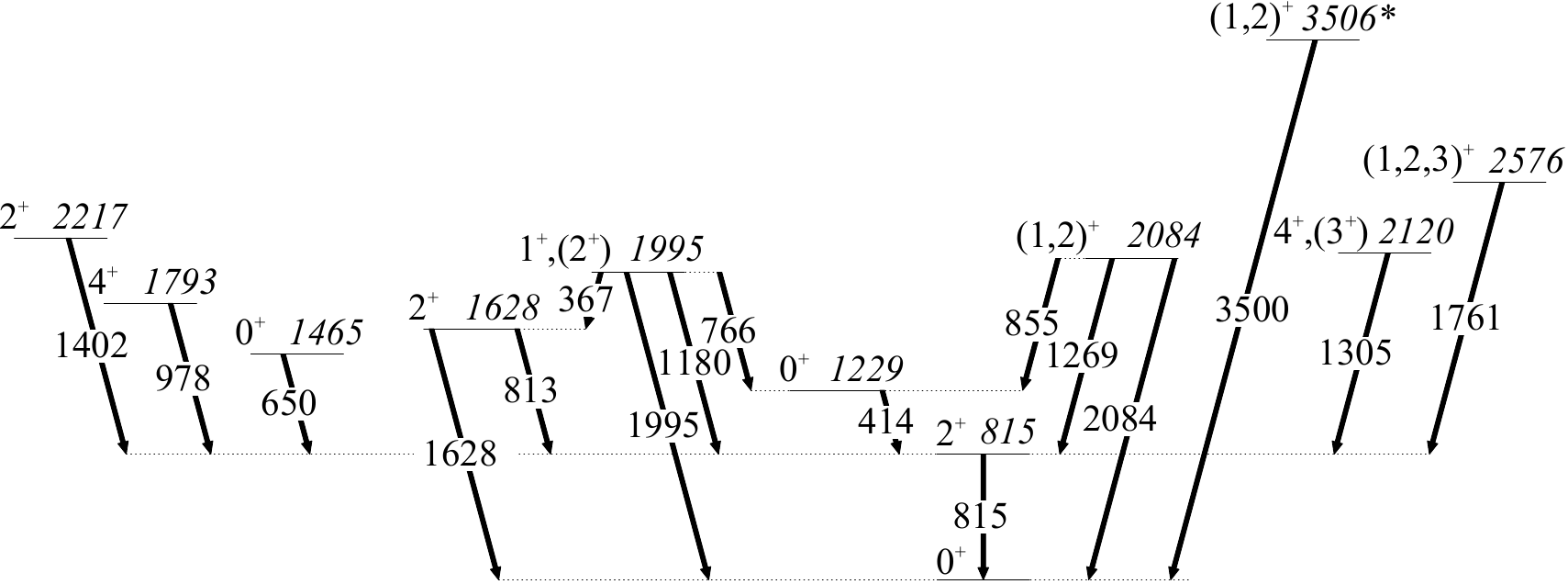}%
\caption{Level scheme of states in \nuc{96}{Sr} that were populated in the \dpsr{95} reaction. The newly observed level at 3506~keV is indicated by a star.}
\label{fig:Levels96}
\end{figure*}

The $\gamma$ rays and excitation energy of states in \nuc{96}{Sr} that were populated via the \dpsr{95} reaction are shown in Fig.~\ref{fig:ExcGam96}.
The very strong 815~keV $\gamma$-ray line visible over the whole excitation energy range indicates that many excited states decay to the 815~keV $2_1^+$ state. An angular distribution analysis was carried out for a total of 10 states in \nuc{96}{Sr}, up to and including a newly observed state at 3506(5)~keV. Substantial population of states above this energy was observed as well, although it was not possible to identify individual states based on the measured $\gamma$ rays. 
Fig.~\ref{fig:Levels96} shows the \nuc{96}{Sr} level scheme for states that were identified in this experiment.

\begin{figure}[h!]
\includegraphics[width=\columnwidth]{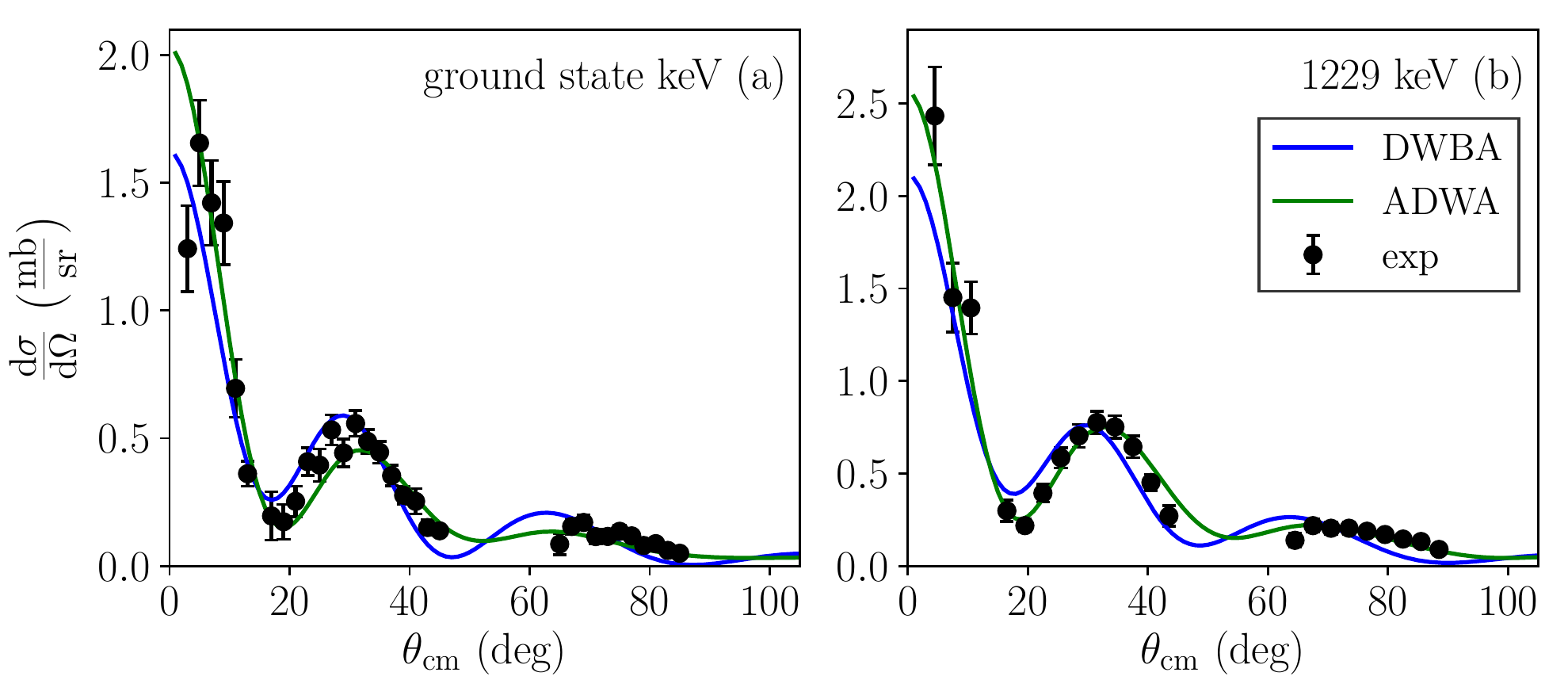}%
\caption{Angular distributions for $\Delta \ell =0$ states in \nuc{96}{Sr}. The experimental data is presented alongside the fitted DWBA (blue) and ADWA (green) calculations, respectively.}
\label{fig:AngDist96_L0}
\end{figure}

\textit{The $0^+$ states:} 
The known 0, 1229 and 1465~keV $0^+$ states were populated in the \dpsr{95} experiment. The main results were already presented in ref.~\cite{cruz18}, here we just summarize the results for the $0^+$ states.
The ground state angular distribution was extracted by fitting the background of the excitation energy spectrum with a constrained exponential function ($\chi^2\sim1$) and taking the excess counts in the range $-0.5<E_\text{x}<0.5$~MeV. 
The 1229~keV $0_2^+$ state angular distribution was produced by gating on the $0_2^+ \rightarrow 2_1^+$ 414~keV $\gamma$ ray. Both angular distributions (Fig.~\ref{fig:AngDist96_L0}) are in very good agreement with the calculated $\Delta \ell=0$ DWBA distributions. The spectroscopic factors for the 0 and 1229~keV \zps\ were determined to be 0.19(3) and 0.22(3), respectively.

For the 1465~keV $0_3^+$ state, it was not possible to extract an angular distribution by gating on the $0_3^+ \rightarrow 2_1^+$ 650~keV $\gamma$ ray owing to its long half-life of $6.7(10)$~ns. The $\gamma$-ray detection efficiency of TIGRESS was simulated using GEANT4~\cite{Geant} for both prompt and isomeric decays from a fast-moving ($\beta = 0.1$) \nuc{96}{Sr} ejectile. The simulations also take into account attenuation of the $\gamma$ rays in the chamber and beam-line materials.
The long half-life of the 1465~keV state results in a large decrease in $\gamma$-ray detection efficiency 
and poor Doppler reconstruction as it was not possible to determine the decay position of \nuc{96}{Sr}. The shape of the Doppler-reconstructed photo-peak was found to depend strongly on the position of the TIGRESS detectors, with clovers positioned at $\theta_\text{lab}>120^\circ$ being the least affected.
 A $\gamma$-ray analysis was used to determine the relative population strengths of the two excited $0^+$ states in \nuc{96}{Sr} by comparing counts in the 414~keV $0^+_2 \rightarrow 2^+_1$ and 650~keV $0^+_3 \rightarrow 2^+_1$ peaks under identical gate conditions. 
 A 1~MeV excitation energy window was used so that both the 1229 and 1465~keV \nuc{96}{Sr} states could be fully included within the energy window, given the resolution of SHARC. This analysis was carried out using only the most downstream TIGRESS detectors positioned at $\theta_\text{lab}>120^\circ$. The ratio of counts in the peaks (after correcting for the relative TIGRESS efficiency) was determined to be 0.22(4). This ratio was compared to the simulation results, which also take into account the indirect feeding of the 1229~keV state from the 1465~keV state via the $0^+_3 \rightarrow 0^+_2$ $E0$ transition and the branching ratio of the 650~keV transition. The experimentally measured relative population strengths are consistent with a scenario where the relative population of the 1465 to the 1229~keV state was 1.50(52). The spectroscopic factor for the 1465~keV state given in Table~\ref{tab:Summary} is this relative population strength ratio multiplied by the 1229~keV state's spectroscopic factor as determined above. The DWBA calculations for both of these states predict the same integrated cross section within $\sim3\%$, and so no excitation energy correction was applied.

\textit{The 815~keV state:} 
It was not possible to extract an angular distribution for this state owing to the weak direct population, strong feeding from the 1229~keV state, and the $E_\text{x}$ resolution. Instead, a $\gamma$-ray analysis was used to estimate the population strength. An energy gate of $0.4<E_\text{x}<1.2$~MeV in the upstream sections of SHARC was used so that all contributions from the 815~keV state were included.
The indirect feeding from the 1229~keV state was subtracted based on the yield of the 414~keV transition, corrected for the TIGRESS efficiency. The 815~keV transition could not be resolved from the close-lying 813~keV transition originating from the 1628~keV state. The known branching ratio of the ground state decay allowed for the determination of the relative population of the 815 and 1628~keV states. The spectroscopic factor for the transfer to the 815~keV state listed in Table~\ref{tab:Summary} was then obtained using this ratio and the result for the 1628~keV state, see below, after correcting for the $Q$-value dependence of the calculated DWBA cross section for transfer to $1d_{3/2}$ neutron orbital.

\begin{figure}[h!]
\includegraphics[width=\columnwidth]{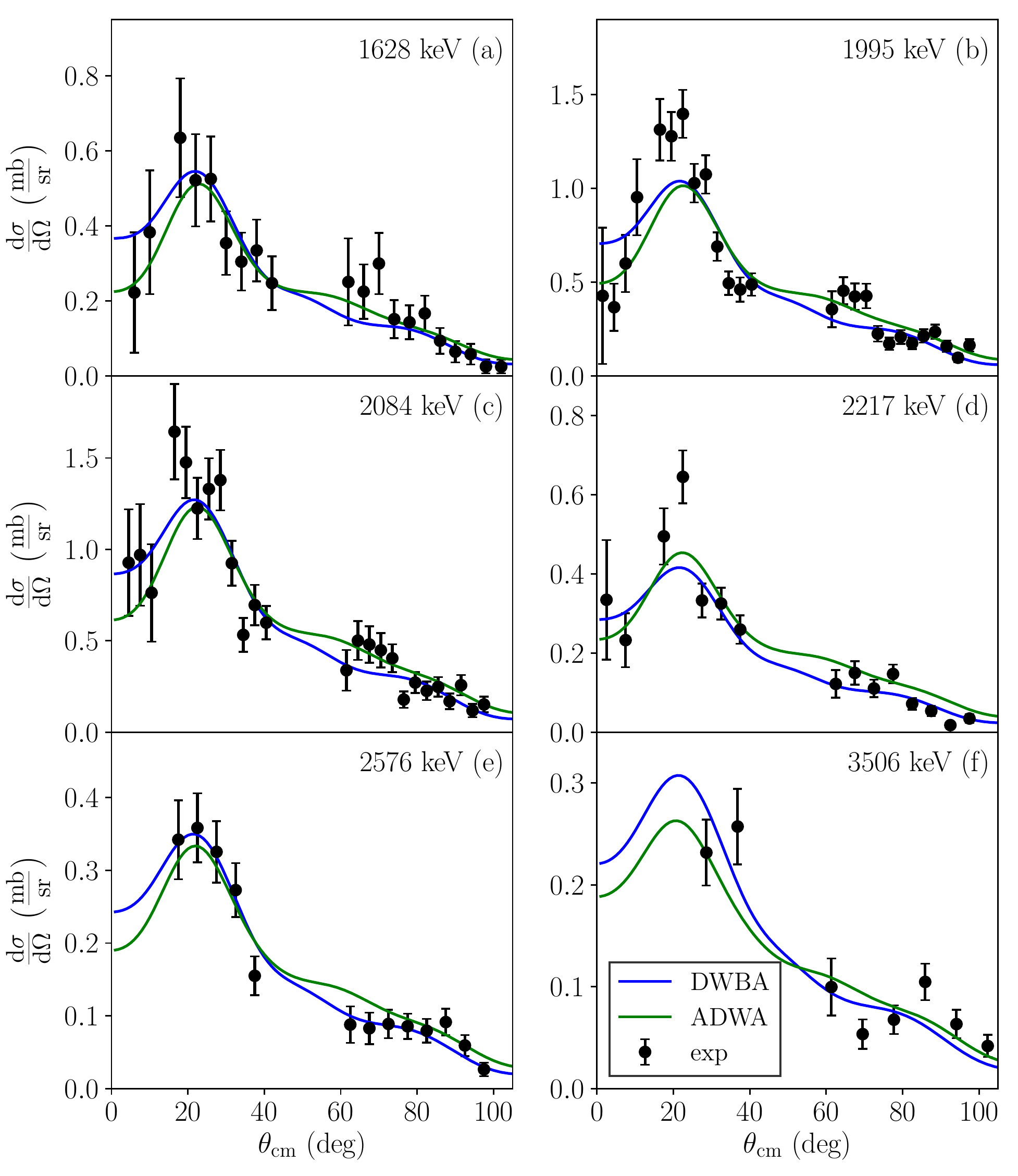}%
\caption{Angular distributions for $\Delta \ell =2$ states in \nuc{96}{Sr}. The experimental data is presented alongside the fitted DWBA (blue) and ADWA (green) calculations, respectively.}
\label{fig:AngDist96_L2}
\end{figure}

\textit{The 1628~keV state:} 
The 1628~keV state decays most strongly to the $2_1^+$ state at 815~keV by the emission of a 813~keV $\gamma$ ray. An angular distribution was thus extracted by double gating on both coincident 813~keV and 815~keV $\gamma$ rays.
The resulting angular distribution, shown in Fig.~\ref{fig:AngDist96_L2} (a), is in very good agreement with the $\Delta \ell=2$ DWBA calculation. This, therefore, constrains the spin and parity to be $1^+$, $2^+$, or $3^+$. 
A suggested spin and parity of $2^+$ was assigned to this state through $\beta$-decay studies of \nuc{96}{Rb}~\cite{Jung1980} using $\gamma$-$\gamma$ angular correlations between the 813~keV and 815~keV transitions, although $1^+$ could not be completely ruled out given the available statistics. Although weak, the branching ratios of this state to the $0^+_{1,2}$ states~\cite{Jung1980} make it highly unlikely that this state has spin and parity $3^+$. If this state were $1^+$, the decay to the $0^+_{1,2}$ states would be of pure $M1$ character. The single-particle Weisskopf estimates for the strength of these $M1$ transitions indicate that they would be similar in strength to the 813~keV transition, but they are measured to be only 12.2 and 5.3\%, respectively. These observations favor a $J^\pi = 2^+$ assignment for the 1628~keV state. The spectroscopic factor listed in Table~\ref{tab:Summary} assumes transfer to the neutron $1d_{3/2}$ orbital, as the $1d_{5/2}$ orbital is considered to be fully occupied at $N=56$.

\begin{figure}[h!]
\includegraphics[width=\columnwidth]{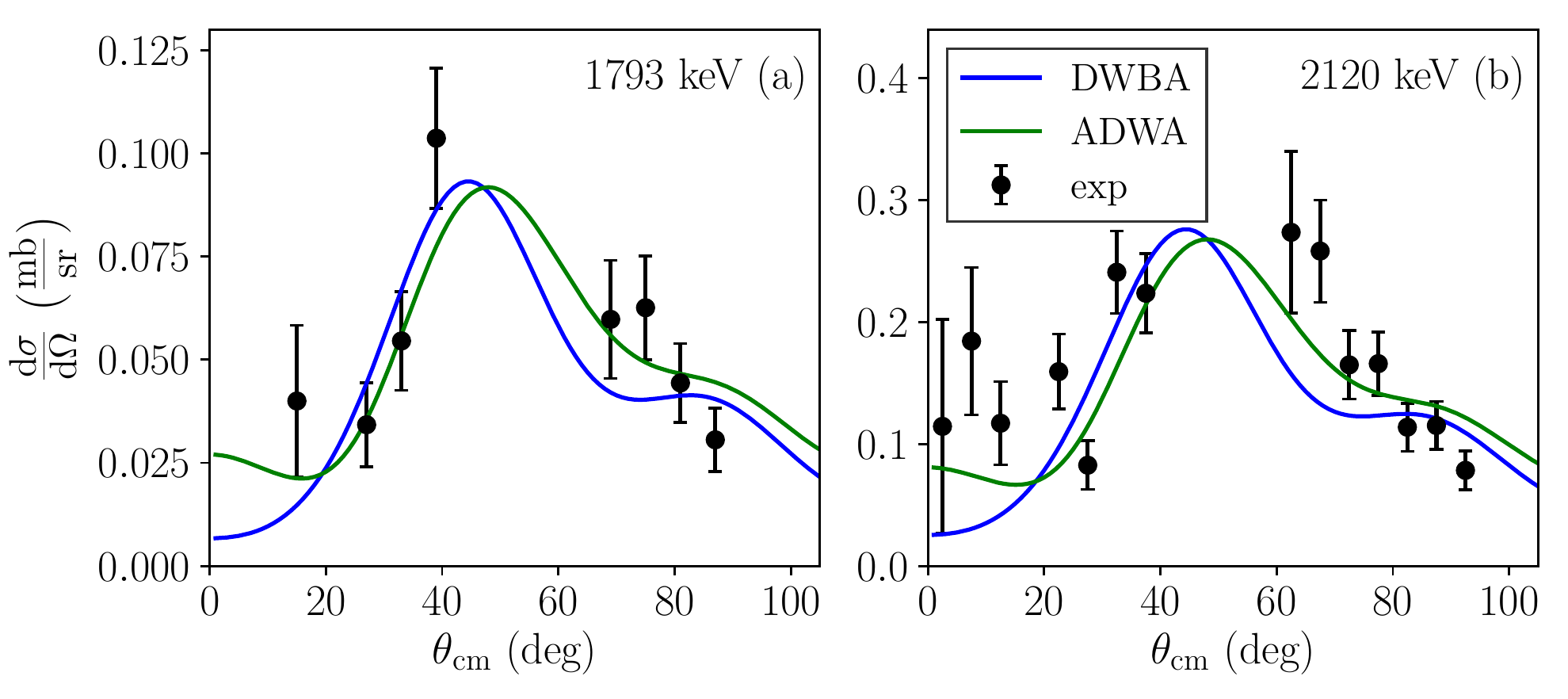}%
\caption{Angular distributions for $\Delta \ell =4$ states in \nuc{96}{Sr}. The experimental data is presented alongside the fitted DWBA (blue) and ADWA (green) calculations, respectively. Potential contamination of the 2120~keV state angular distribution by the neighboring 2113~keV state has been neglected (see text).}
\label{fig:AngDist96_L4}
\end{figure}

\textit{The 1793~keV state:} 
This state was weakly populated, with most of the observed $\gamma$-ray strength coming from indirect feeding from higher levels.
Fig.~\ref{fig:AngDist96_L4} (a) shows the angular distribution for the 1793~keV state, which was produced by gating on the $4_1^+ \rightarrow 2_1^+$ 978~keV $\gamma$ ray transition. The measured angular distribution, which was best reproduced by a $\Delta \ell=4$ DWBA calculation, is consistent with the established spin of $4^+$~\cite{Jung1980}.

\textit{The 1995~keV state:} 
This state was strongly populated directly through the \dpsr{95} transfer reaction, with negligible indirect feeding. 
It can be clearly seen in Fig.~\ref{fig:ExcGam96} as a strong 1180~keV $\gamma$ ray in coincidence with excitation energies in the range $1.6<E_\text{x}<2.4$~MeV.
The angular distribution, shown in Fig.~\ref{fig:AngDist96_L2} (b) was produced by gating on the 1180~keV $\gamma$ ray. It shows clear $\Delta \ell=2$ character which constrains the spin and parity to be $1^+$, $2^+$, or $3^+$. A spin and parity of $3^+$ is unlikely since decay to the ground and $0^+_2$ states has been observed. A $J^\pi = 1^+$ assignment was suggested based on $\beta$-decay studies of \nuc{96}{Rb}~\cite{Jung1980} using $\gamma$-$\gamma$ angular correlations between the 1180~keV and 815~keV $\gamma$ rays. For completeness, Table~\ref{tab:Summary} also lists the $1d_{3/2}$ spectroscopic factor for the $J^\pi = 2^+$ assignment.

\textit{The 2084~keV state:} 
This state was also strongly populated with negligible feeding from higher lying states. The direct ground state decay can be clearly seen in Fig.~\ref{fig:ExcGam96} as a strong 2084~keV $\gamma$-ray line in coincidence with excitation energies in the range $1.6<E_\text{x}<2.4$~MeV. The angular distribution obtained by gating on this transition (Fig.~\ref{fig:AngDist96_L2} (c)) shows clear a $\Delta \ell=2$ character constraining the spin and parity of this state to $1^+,2^+$ or $3^+$. Using similar arguments as for the 1995~keV level, the decay branches to the $0_{1,2}^+$ states effectively rule out $3^+$.
 The $\log \mathit{ft}$ value of the $\beta$-decay of the \nuc{96}{Rb} $2^{(-)}$ ground state to the 2084~keV state suggests a first forbidden transition which, together with the present result, constrains this state to have spin and parity $1^+$ or $2^+$. 

\textit{The 2120~keV state:} 
The main (91~\%) decay branch of this state is by a 1305~keV transition to the $2^+$ state. However, it cannot be resolved from the 1299~keV transition arising from the 2113~keV state given the TIGRESS energy resolution after Doppler-correction. The 2113~keV state also decays by 485~keV (branching ratio 22~\%) and 607~keV (35\%) $\gamma$ rays which have been observed in the excitation energy range $1.8<E_\text{x}<2.6$~MeV. This indicates that the relative population strengths are 25(20)\% for the 2113~keV level and 75(20)\% for the 2120~keV state. 
The angular distribution gated on both the 1299 and 1305~keV $\gamma$-ray lines  shown in Fig.~\ref{fig:AngDist96_L4} (b) is thus dominated by the 2120~keV state. It is in best agreement with $\Delta \ell =4$ which is in accord with the tentative assignment $J=4$ from spontaneous fission studies of \nuc{248}{Cm}~\cite{Wu2004}.
The spectroscopic factor for transfer to the $0g_{7/2}$ orbital given in Table~\ref{tab:Summary} is an upper limit for the 2120~keV state ignoring the contribution of the 2113~keV level to the angular distribution. 

\textit{The 2217~keV state:} 
The angular distribution shown in Fig.~\ref{fig:AngDist96_L2} (d)  was produced by gating on the 1402~keV $\gamma$-ray transition depopulating this state and is well described by a $\Delta \ell=2$ calculation. Therefore $J^\pi=2^+$ is assigned to this state confirming the previous provisional $J=2$ assignment based on $\gamma$-$\gamma$ angular correlation measurements~\cite{Jung1980}.

\textit{The 2576~keV state:} 
The angular distribution for this level (Fig.~\ref{fig:AngDist96_L2} (e)) was produced by gating on the 1761~keV $\gamma$-ray transition. It has previously been observed only in $\beta$-decay of \nuc{96}{Rb}~\cite{ENSDF} and its strength suggests a first-forbidden decay. 
This is in agreement with the $\Delta \ell = 2$ angular distribution deduced here, which constrains the spin and parity to be $1^+$, $2^+$ or $3^+$.
Spectroscopic factors assuming transfer to the $1d_{3/2}$ ($0g_{7/2}$) neutron orbital for $J^\pi = 1^+,2^+$ ($3^+$) are listed in Table~\ref{tab:Summary}.

\textit{The 3506~keV state:} 
The $3506(6)$~keV transition is newly observed in this work (inset of Fig.~\ref{fig:ExcGam96}). The excitation energy spectrum gated on this transition shows that this is a direct ground state decay. The angular distribution obtained by gating on this $\gamma$ ray is shown in Fig.~\ref{fig:AngDist96_L2} (f). The measured angular distribution is in good agreement with the $\Delta \ell=2$ DWBA calculation. No other new or known transitions were observed when gating on this excitation energy range, indicating that the branching ratio for the 3506~keV $\gamma$ ray to the ground state is 100(10)\%. This constrains the spin and parity to be $1^+$ or $2^+$.

\subsection{The \dpsr{96} reaction}
\begin{figure}[h!]
\includegraphics[width=\columnwidth]{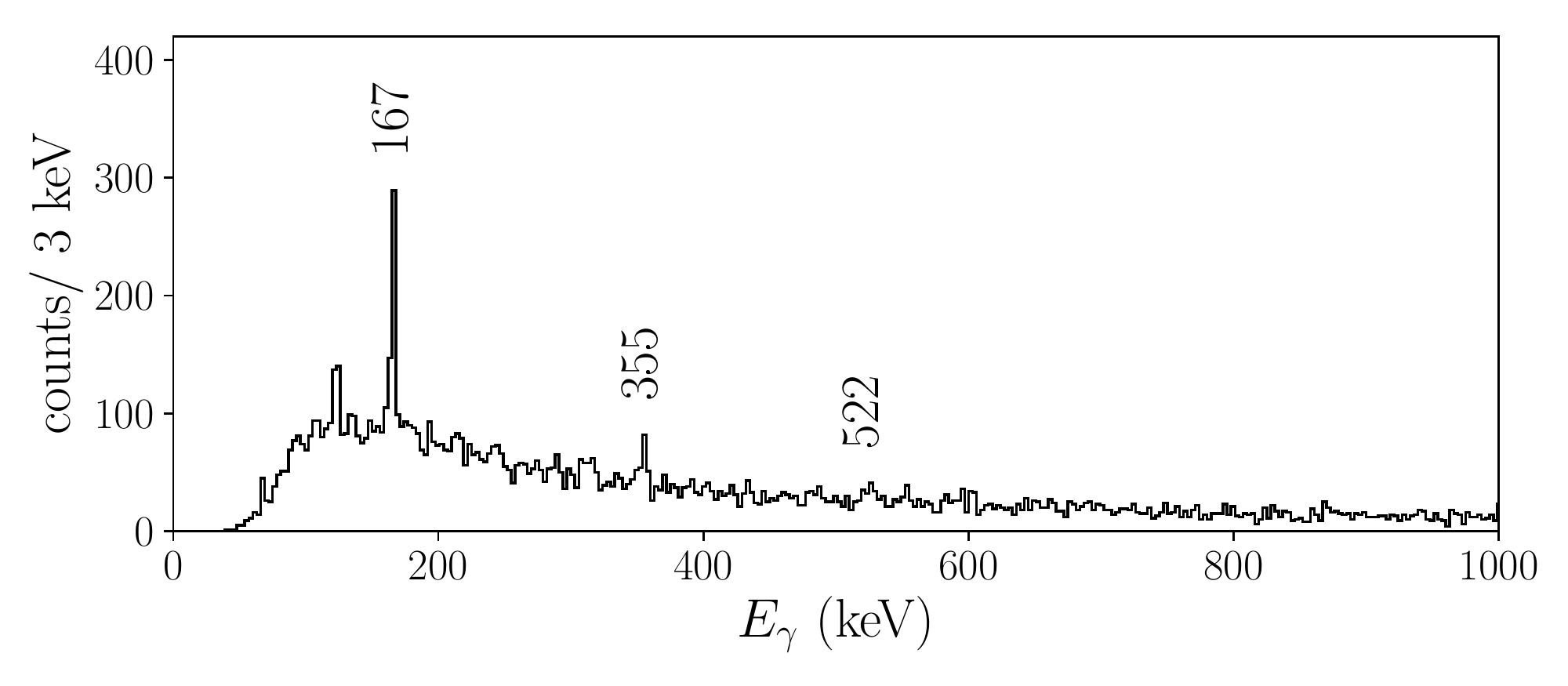}%
\caption{Projected $\gamma$-ray spectrum for \nuc{97}{Sr} states populated via the \dpsr{96} reaction. A cut on excitation energies below 1~MeV has been applied. 
}
\label{fig:ExcGam97}
\end{figure}

\begin{figure}[h!]
\includegraphics[height=0.85\columnwidth,angle=-90]{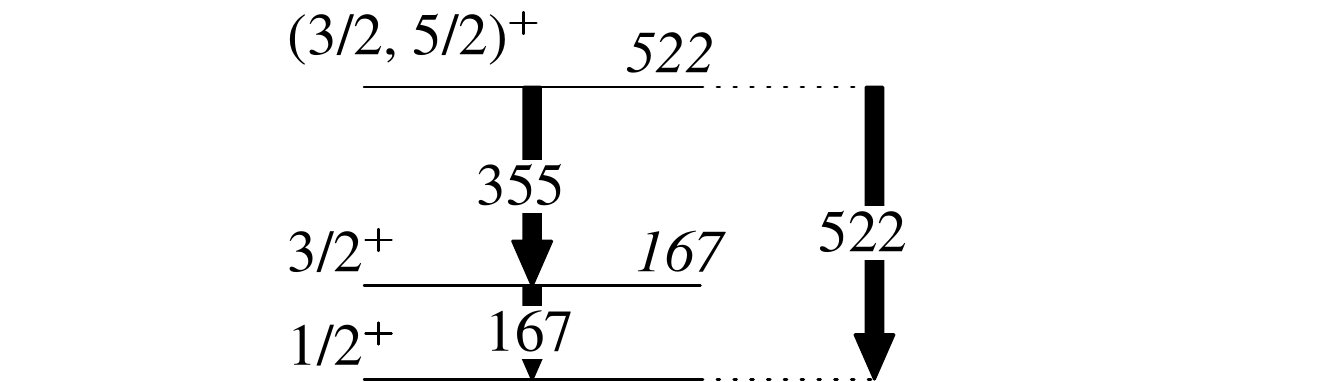}%
\caption{Level scheme for \nuc{97}{Sr} states that were populated through \dpsr{96}.}
\label{fig:Levels97}
\end{figure}

\begin{figure}[h]
\includegraphics[width=\columnwidth]{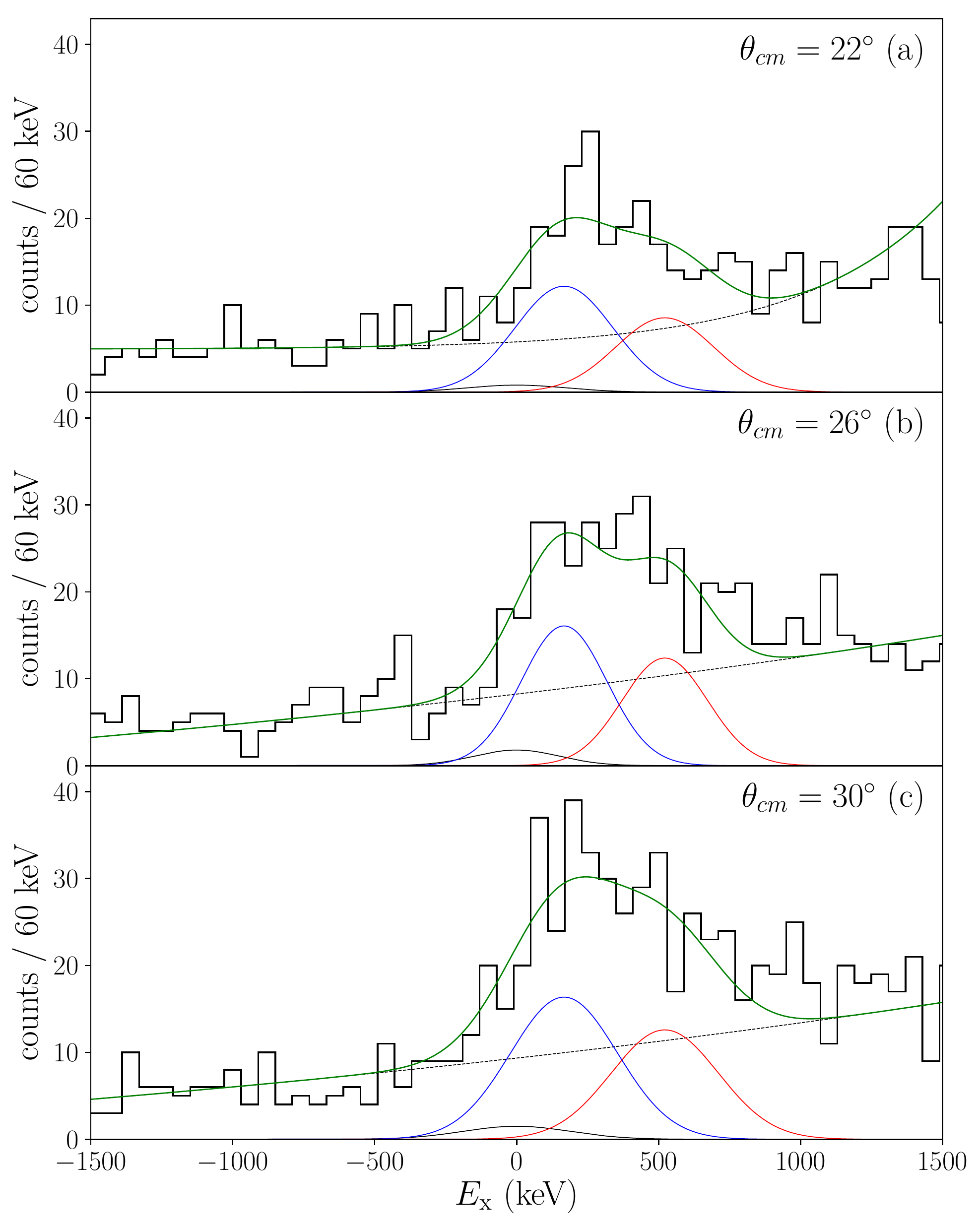}%
\caption{Excitation energy spectrum extracted from the recoiling proton energies and angles at a center of mass angles $\theta_\text{cm} = 22, 26$, and $30^\circ$. The continuous green line shows the constrained 3-peak fit of the 0, 167 and 522~keV states. The dashed line represents the continuous background.}
\label{fig:3PeakFit_96}
\end{figure}

The $\gamma$ rays and excitation energy of states in \nuc{97}{Sr} that were populated via the \dpsr{96} reaction are shown in Fig.~\ref{fig:ExcGam97}.
The 167 and 355~keV $\gamma$ rays in the energy range $-0.5<E_\text{x}<1$~MeV indicate that both the known 167 and 522~keV excited states were populated in this experiment. Fig.~\ref{fig:Levels97} shows the \nuc{97}{Sr} level scheme for states that were identified in this work. 
No other excited states could be unambiguously identified, owing to the limited statistics.
Given the small difference in energy between the ground state and 167~keV first excited state, and the $E_\text{x}$ energy resolution, it was not possible to obtain the cross sections and angular distributions based on the excitation energy spectrum alone. 
The strength of the ground state was thus derived by means of a constrained three-peak fit for the 0, 167 and 522~keV states as discussed above for \nuc{95}{Sr}. Examples are shown in Fig.~\ref{fig:3PeakFit_96}.

\begin{figure}[h!]
\includegraphics[width=\columnwidth]{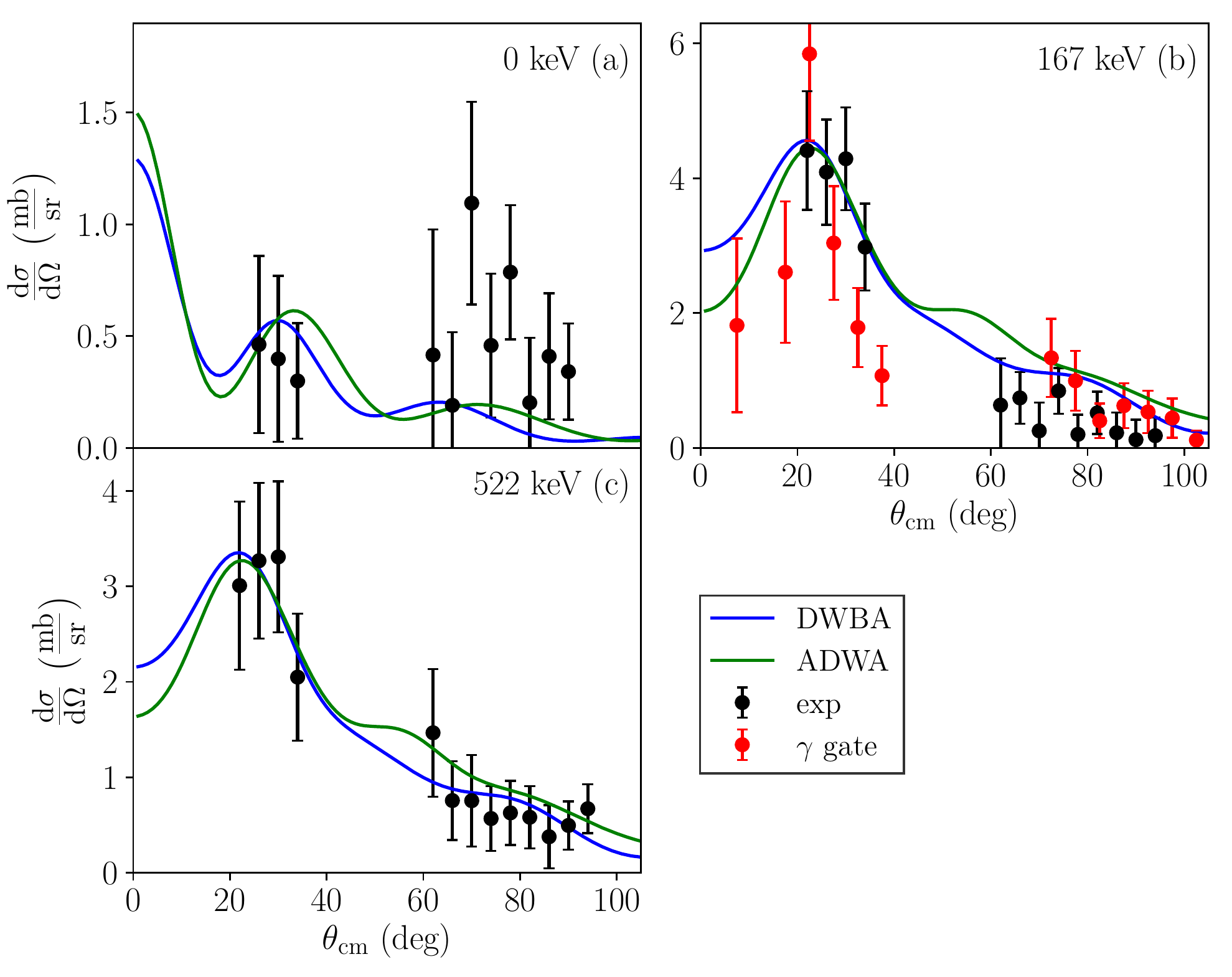}%
\caption{Fit of the reaction model calculations to the experimental data for the 167 and 522~keV states in \nuc{97}{Sr}. The solid lines are the best-fitting reaction model calculations using the DWBA (blue) and ADWA (green) methods. The fitting was restricted to the forward angles ($\theta_\text{cm} < 40^\circ$). For the 167~keV state the angular distribution extracted by gating on the 167~keV $\gamma$-ray transition and the excitation energy is also shown.}
\label{fig:AngDist97}
\end{figure}

\textit{The ground state:} 
The ground state was very weakly populated through the \dpsr{96} reaction and the angular distribution shown in Fig.~\ref{fig:AngDist97} (a) did not exhibit a clear shape as no data could be obtained for the smallest scattering angles ($\theta_\text{cm} < 20^\circ$). In this region the yield is expected to be very small and due to the small $Q$-value the background is high at low excitation energy.
However, the ground state is known to be $J^\pi =1/2^+$~\cite{Buchinger1990} and the angular distribution obtained is in accord with $\Delta \ell =0$. The spectroscopic factor given in Table~\ref{tab:Summary} has been extracted from the data shown in Fig.~\ref{fig:AngDist97} (a) as well as a two-component fit of the summed angular distributions of the ground and 167~keV states. 

\textit{The 167~keV state:} 
Two independent angular distributions were produced for the 167~keV state; one was extracted using the three peak fit (Fig.~\ref{fig:AngDist97} (b)) and a second was derived by gating on the 167~keV $\gamma$ ray and the excitation energy limiting the contribution from the 522~keV state.
The shape of both angular distributions are in good agreement with the $\Delta \ell=2$ reaction model calculations, in agreement with the established spin and parity of $3/2^+$~\cite{Kratz1983}. The spectroscopic factors that were extracted for each of the methods are 0.25(7) and 0.24(8), respectively, assuming the addition of a neutron to the $1d_{3/2}$ orbital. The weighted average of the two spectroscopic factors is given in Table~\ref{tab:Summary}. 

\textit{The 522~keV state:} 
The small number of counts in the 355 and 522~keV $\gamma$-ray peaks (shown in Fig.~\ref{fig:ExcGam97}) did not allow for a $\gamma$-gated angular distribution for the 522~keV state, and so the spectroscopic factor for this state was determined by using the three-peak fit. 
The $\Delta\ell = 2$ angular distribution shown in Fig.~\ref{fig:AngDist97} (c) constrains the $J^\pi$ of this state to be $3/2^+$ or $5/2^+$, in agreement with the $M1$ multipolarities of the decay to the 167~keV state and also from the 687~keV $5/2^+$ state~\cite{Kratz1983}.
The population of this state by adding a neutron to the $1d_{3/2}$ orbital is most likely as the $1d_{5/2}$ orbital is expected to be fully occupied at $N=59$ and the spectroscopic factor should be even lower than in \nuc{95}{Sr}. Consequently, $3/2^+$ is a more likely spin and parity for this state.
For completeness, Table~\ref{tab:Summary} includes the spectroscopic factors for both possibilities 0.21(8) and 0.13(5) for $J^\pi = 3/2^+$ and $5/2^+$, respectively, using the DWBA calculations.
\begin{table*}[t!]           
\begin{center}
\def\arraystretch{1.3} \setlength\tabcolsep{10pt}
  \begin{tabular}{ccccccc}
    \hline
    \hline    
    Nucleus & $E_\text{x}$ [keV] & $E_\gamma$ [keV] & $J^\pi$ & $\Delta\ell$ & $C^2S$ (DWBA)  & $C^2S$ (ADWA) \\
    \hline
    \nuc{95}{Sr} & 0     &  fit  	   & $\frac{1}{2}^+$                                       & 0 & 0.41(9)                       & 0.34(7) \\ 
    &		   352   &  fit, 352 	   & $\mathbf{\frac{3}{2}^+}$                              & 2 & 0.53(8)$^\dag$                & 0.45(7)$^\dag$ \\    
    &		   556   &  -	 	   & $\mathbf{\frac{7}{2}^+}$                              & - & -                             & - \\        
    &		   681   &  fit, 329, 681  & $\mathbf{\frac{5}{2}^+}$                              & 2 & 0.16(3)$^\dag$                & 0.14(3)$^\dag$ \\   
    &		   1239  &  -	 	   & $\mathbf{\frac{3}{2}^+,\frac{5}{2}^+,\frac{7}{2}^+}$  & - & -                             & - \\        
    &		   1666  &  -	 	   & $\mathbf{\frac{3}{2}^+,\frac{5}{2}^+,\frac{7}{2}^+}$  & - & -                             & - \\              
    \hline                                      
    \nuc{96}{Sr} & 0    &  fit		   & $0^+$  		                                   & 0 & 0.19(3)                       & 0.15(3)  \\  
    &		   815  & -		   & $2^+$  		                                   & - & 0.038(12)                     & 0.034(12) \\ 
    &		   1229 &  414		   & $0^+$  		                                   & 0 & 0.22(3)                       & 0.19(3)  \\ 
    &		   1465 &  -		   & $0^+$  		                                   & - & 0.33(13)                      & 0.29(12) \\ 
    &		   1628 &  813 $+$ 815	   & $\mathbf{2^+}$	                                   & 2 & 0.069(25)                     & 0.056(23) \\ 
    &		   1793 &  978		   & $4^+$  		                                   & 4 & 0.066(16)                     & 0.058(17) \\ 
    &		   1995 &  1180		   & $1^+, (2^+)$                                          & 2 & 0.20(3), (0.12(2))            & 0.18(3), (0.10(2))  \\     
    &		   2084 &  2084		   & $1^+,2^+$ 	                                           & 2 & 0.24(5), 0.15(3)              & 0.21(4), 0.12(3)  \\     
    &		   2120 &  1305		   & $4^+, (3^+)$ 	                                   & 4 & 0.19(4), (0.21(4))            & 0.16(4), (0.21(4))   \\ 
    &		   2217 &  1402		   & $2^+$  		                                   & 2 & 0.047(8)                      & 0.034(8)  \\ 
    &		   2576 &  1761		   & $\mathbf{1^+,2^+,3^+}$                                & 2 & 0.062(12), 0.037(7),          & 0.049(9),0.028(6), \\
    &		        &  		   &                                                       &   &                      0.025(5) &                   0.019(5)  \\ 
    &		   3506(5)$^*$ &  3506(5)	   & $\mathbf{1^+,2^+}$  	                           & 2 & 0.047(9), 0.027(5)            & 0.034(8), 0.020(4)  \\  
   \hline                                       
    \nuc{97}{Sr} & 0     &  fit     	   & $\frac{1}{2}^+$                                       & 0 & 0.07(5)                       & 0.06(5) \\ 
                 &       &                 &                                                       &   & 0.11(10)$^\ddagger$           & 0.07(7)$^\ddagger$\\
    &		   167   &  fit, 167 	   & $\frac{3}{2}^+$                                       & 2 & 0.25(5)$^\dag$                & 0.20(5)$^\dag$ \\    
                 &       &                 &                                                       &   & 0.21(7)$^\ddagger$            & 0.19(7)$^\ddagger$\\
    &		   522   &  fit		   & $\frac{3}{2}^+,\frac{5}{2}^+$                         & 2 & 0.21(8), 0.13(5)              & 0.17(7), 0.11(4) \\       
   \hline
    \hline
    \multicolumn{7}{l}{$^\dag$\footnotesize{$C^2S$ presented is the weighted average from multiple determinations}}\\
    \multicolumn{7}{l}{$^*$\footnotesize{new state}}\\
    \multicolumn{7}{l}{$^\ddagger$\footnotesize{determined from the summed angular distribution of ground and 167~keV state}}\\
  \end{tabular} 
  \caption{Results for \nuc{95,96,97}{Sr} states that were studied through the \dpsr{94,95,96} reactions. 
Spectroscopic factors ($C^2S$) are given for all allowed $J^\pi$. $J^\pi$ values in bold are new assignments or refined constraints.  
 The method of angular distribution extraction, if any, for each state is presented under $E_\gamma$. Assignments and spectroscopic factors in parenthesis are alternative assignments that cannot be definitively ruled out by the present data, but are unlikely given previous experiments. }
  \label{tab:Summary}
\end{center}
\end{table*}

\section{Discussion}
The results obtained here can be used to gain insights into the underlying single-particle configurations of states in \nuc{95,96,97}{Sr}. 
The results are compared in the following to shell model calculations to investigate the role of proton and neutron configurations in the low-lying states. While the present calculations are not well adapted to describe the deformed structures in \nuc{96}{Sr} and \nuc{97}{Sr}, the structure of \nuc{95}{Sr} before the shape transition should be well described, even in rather limited model spaces as will be discussed.
 
Shell model calculations for \nuc{94-97}{Sr} were carried out using NushellX~\cite{NushellX} with the \textit{glek} interaction~\cite{GLEK} and several different model spaces. The single-particle energies of the interaction were adjusted so that the energies of low-lying states in the vicinity of $N\sim56$ and $Z\sim38$ were in good agreement with experiment~\cite{cruz18}.
In the present calculations the neutron $1d_{5/2}$, $2s_{1/2}$ $1d_{3/2}$ and $0g_{7/2}$ orbitals, outside an inert $N = 50$ core, were included. The higher-lying $0h_{11/2}$ orbital was not included as contributions from this orbital to low-lying positive parity states are expected to be small owing to the high single-particle energy~\cite{Sieja2009}.

Three different truncations of the proton valence space were investigated. In the smallest model space (a) the protons were frozen in a $(1p_{3/2})^4$ configuration so that the calculated states were built up using only the neutron configurations. 
Model space (b) included the $1p_{1/2}$ orbital and protons could be distributed across the $1p$ orbitals so that the effect of $(1p_{3/2})^{(4-x)}(1p_{1/2})^x$ configurations could be investigated. 
A third model space, (c), was used to investigate the effect of the proton $0g_{9/2}$ orbital on low-lying states. Up to two protons were allowed to occupy this orbital, so that configurations such as $(1p_{3/2})^2(0g_{9/2})^2$ and $(1p_{1/2})^2(0g_{9/2})^2$  were possible. This truncation was necessary due to the available computational resources.
Proton seniority $\nu \neq0$ configurations are expected to play a negligible role in the configurations of states that are strongly populated via the \dpsr{} reactions as single-step neutron transfer cannot break and re-couple proton pairs. Overall, additional proton degrees of freedom resulted in a lowering of the excitation energies, as correlations between complex configurations provide extra binding energy.
This effect was evidenced by the increased mixing of the large number of configurations in the wave functions. The increased proton model space also impacted the predicted spectroscopic factors, as the mixed wave functions, unsurprisingly, tend to have smaller overlaps.

\subsection{\nuc{95}{Sr}}
In a shell model picture, low-lying states in \nuc{95}{Sr} can be approximated as simple excitations of the unpaired neutron into the different valence orbitals, which define the spins and parities of the low-lying states. 
The ground state spectroscopic factor (Table~\ref{tab:ShellModel9597}) is in good agreement with that calculated in the shell model for all three model spaces, although the substantial improvement in (b) indicates that proton pair excitations into the $1p_{1/2}$ orbital play an important role in the ground states of both \nuc{94}{Sr} and \nuc{95}{Sr}.
\begin{table*}[t!]           
\begin{center}
\def\arraystretch{1.3} \setlength\tabcolsep{10pt}
  \begin{tabular}{cc|cc|cc|cc|cc}
    \hline
    \hline    
    \multicolumn{2}{c|}{} & \multicolumn{2}{c|}{exp.} & \multicolumn{2}{c|}{SM (a)} & \multicolumn{2}{c|}{SM (b)} & \multicolumn{2}{c}{SM (c)} \\
    Nucleus & $J^\pi$  & $E$ (keV) & $C^2S$ &  $E$ (keV) & $C^2S$  & $E$ (keV)  & $C^2S$ & $E$ (keV) & $C^2S$\\		
    \hline
    \nuc{95}{Sr} & $\frac{1}{2}^+$ &    0	& 0.41(9)   &    0 & 0.553  &    0 & 0.449 &    0 & 0.413 \\
    &		   $\frac{3}{2}^+$ &  352	& 0.53(8)   &  766 & 0.865  &  412 & 0.767 &  375 & 0.744 \\   
    &		   $\frac{5}{2}^+$ &  681	& 0.16(3)   &  691 & 0.146  &  585 & 0.180 &  523 & 0.201 \\
    &		   $\frac{7}{2}^+$ &  556	&           & 1086 & 0.959  &  602 & 0.828 &  205 & 0.757 \\
   \hline                                	 
    \nuc{97}{Sr} & $\frac{1}{2}^+$ &    0	&  0.10(5)  & 1631 & 0.013  & 1279 & 0.024 &  417 & 0.002 \\    
    &	           $\frac{3}{2}^+$ &  167	&  0.25(5)  &    0 & 0.881  &    0 & 0.804 &  117 & 0.713 \\    
    &		   $\frac{7}{2}^+$ &  308 	&           &  270 & 0.979  &  149 & 0.931 &    0 & 0.819 \\       
    &		   $\frac{5}{2}^+$ &  522	&  0.13(5)  & 1714 & 0.025  & 1336 & 0.042 &   57 & 0.000 \\       
   \hline
   \hline
  \end{tabular} 
  \caption{Comparison of \dpsr{94,96} spectroscopic factors to shell model calculations for low-lying states. 
    The labels SM (a), (b) and (c) denote the three proton model spaces that were investigated (see text).   }
  \label{tab:ShellModel9597}
\end{center}
\end{table*}
The same is also true for the energy and spectroscopic factor of the $3/2^+$ first excited state: the calculated energy of this level drops substantially with the inclusion of the proton $1p_{1/2}$ orbital. As can be seen, a gradual reduction in spectroscopic strength is predicted for the ground state and 352~keV excited states as the proton degrees of freedom are increased. In each case, there were no other $1/2^+$ or $3/2^+$ states with substantial spectroscopic strength ($C^2S>0.04$) predicted.
On the other hand, each calculation predicted a low-energy $5/2^+$ state with $C^2S>0.15$ at around $\sim600$~keV (Table~\ref{tab:ShellModel9597}) which is dominated by a neutron $(1d_{5/2})^5(2s_{1/2})^2$ configuration in all of the calculations.
The population of such a state in the one-neutron transfer suggests that the $\nu1d_{5/2}$ orbit is not fully occupied in the ground state of \nuc{94}{Sr}. The larger model spaces, which increase the neutron particle-hole configurations in the \nuc{94}{Sr} ground state, show an increase in the spectroscopic factor for the $5/2^+$ state. This also affirms the assignment of $5/2^+$ to the state seen at 681~keV. The spectroscopic factor and the excitation energy of the $7/2^+$ state strongly depends on the proton configurations. This demonstrates the effect of the Federman-Pittel mechanism~\cite{FP1977,FP1979} whereby the mutual interaction of the $\pi 0g_{9/2}$ and $\nu 0g_{7/2}$ orbitals drives the deformation in this region.
While the spectroscopic factor for this state could not be deduced, the observed yield (Fig.~\ref{fig:3PeakFit}) suggests that this state has a small spectroscopic factor, at variance with the shell model calculations.

\begin{figure}[h!]
\includegraphics[width=\columnwidth]{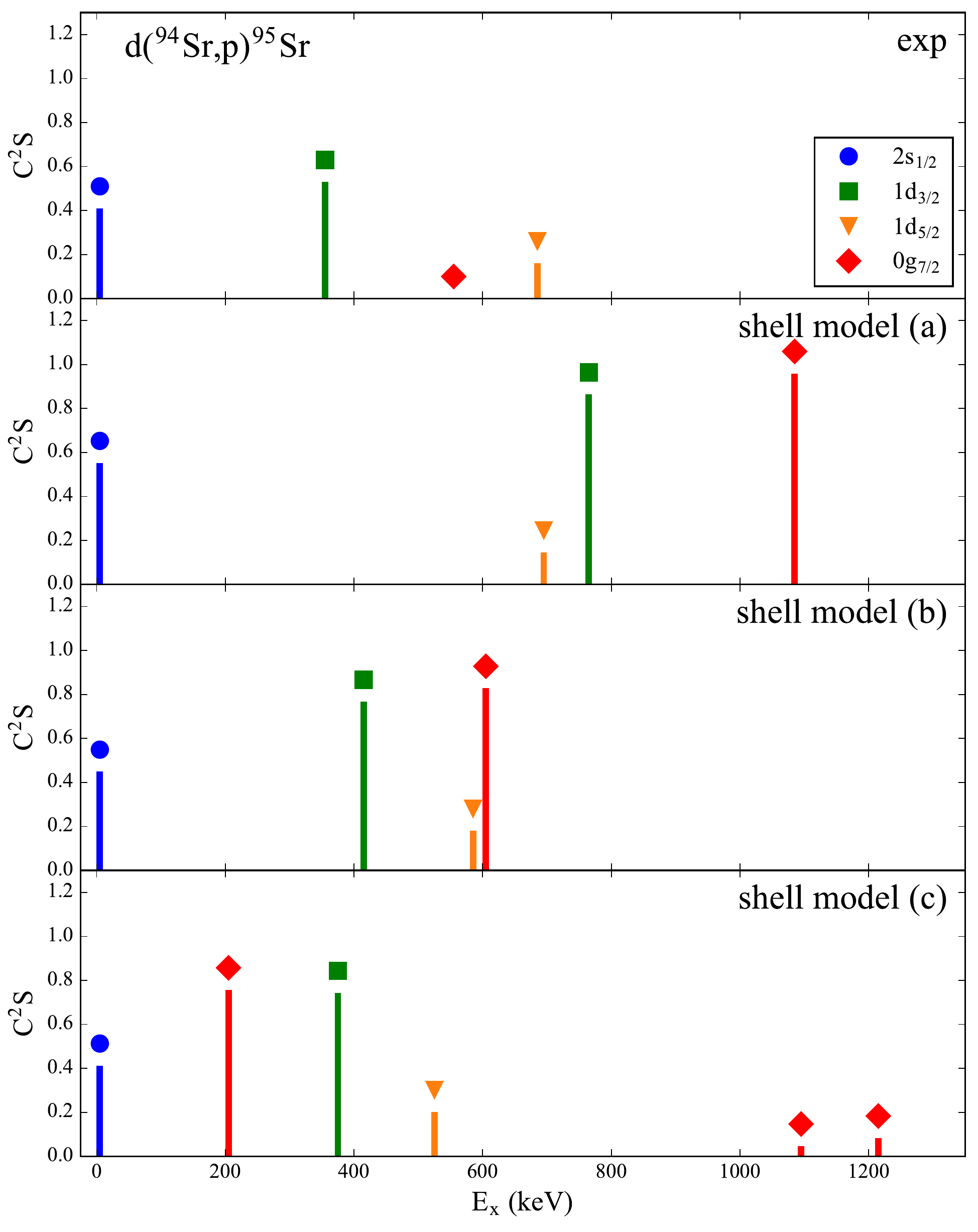}%
\caption{Comparison of experimental (exp) spectroscopic factors ($C^2S$) to those from shell model calculations carried out in model spaces (a), (b) and  (c) -- see text. States are labeled by the neutron single-particle orbital populated in the transfer reaction.}
\label{lev:ShellModel95}
\end{figure}

Figure~\ref{lev:ShellModel95} shows the experimental level energies and DWBA spectroscopic factors for \nuc{95}{Sr} states that were populated via the \dpsr{94} reaction compared to the shell model calculations.
Overall, the shell model calculations for proton model space (b) describe these low-lying states very well aside from the $7/2^+$ state. This suggests that the ground states of both \nuc{94}{Sr} and \nuc{95}{Sr} have similar and nearly spherical shapes and in agreement with $B(E2)$~\cite{Mach1991,Chester2017} and charge radii~\cite{Buchinger1990} measurements.
It should be noted that a recent Monte-Carlo shell model calculation~\cite{Togashi2016} predicts the onset of deformation in the Sr nuclei too early. This is evident from the calculated spectra of the even-even Sr nuclei~\cite{Regis2017} as well as the level scheme of \nuc{95}{Sr} with 13 states below 1~MeV, some of them strongly deformed~\cite{Togashi2018}.

\subsection{\nuc{96}{Sr}}
\begin{figure}[h!]
\includegraphics[width=\columnwidth]{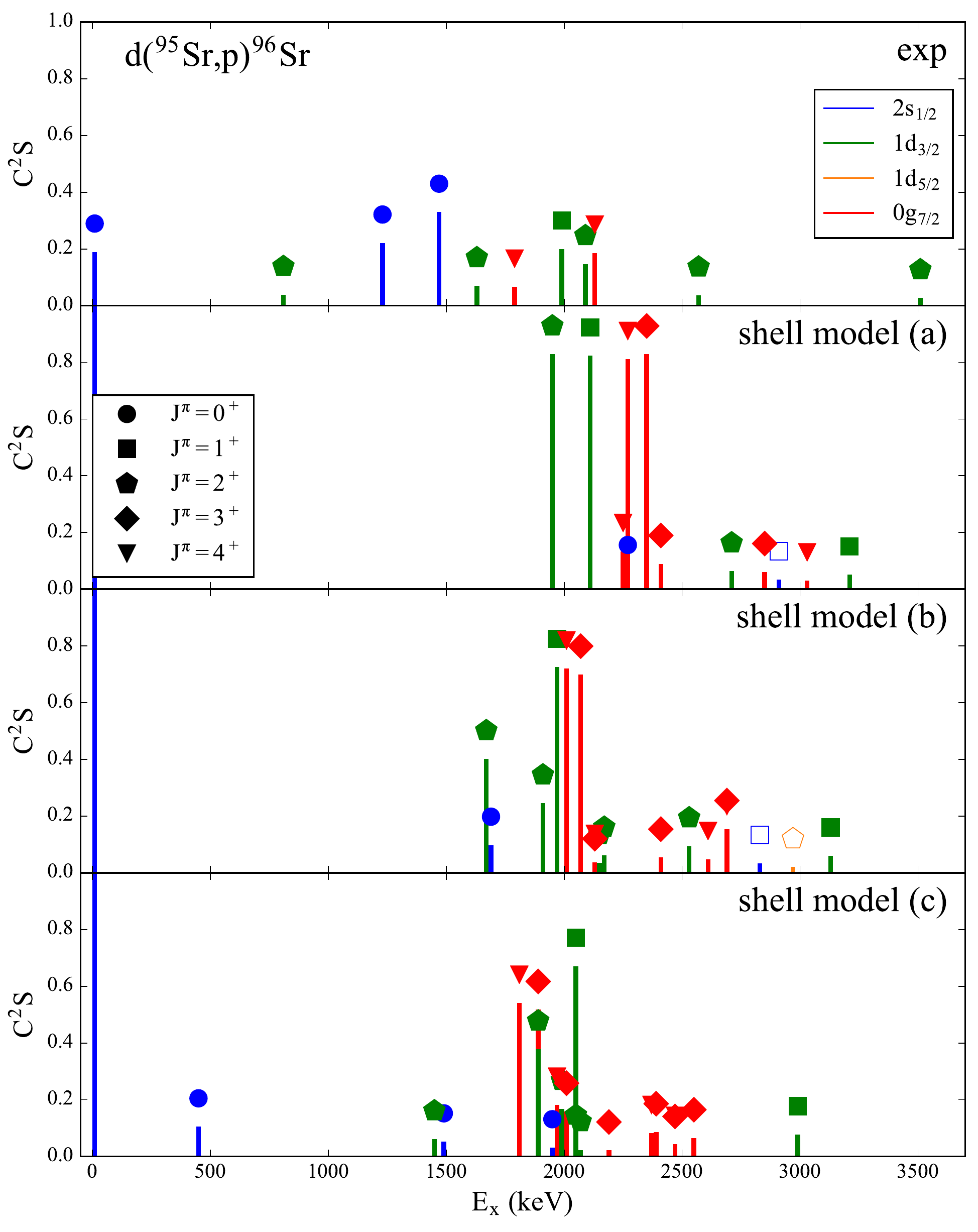}%
\caption{Comparison of experimental (exp) spectroscopic factors ($C^2S$) for \dpsrf{94}{95} to shell model calculations that were carried out in model spaces (a), (b) and  (c) -- see text. States are labeled by their spin and parity as well as the orbital populated in the transfer reaction. Open symbols label the $1^+$ states populated by transfer to the $2s_{1/2}$ orbital, as well as transfer to the $1d_{5/2}$ orbital for $J^\pi = 2,3^+$. 
Only states with $C^2S>0.01$ are shown. For experiment $J^\pi=2^+$ has been assumed for the 2084, 2576, and 3506~keV.
}
\label{fig:ShellModel96}
\end{figure}

Table~\ref{tab:ShellModel96} compares the shell model results within each proton model space for the lowest states.
\begin{table*}[t!]           
\begin{center}
\def\arraystretch{1.3} \setlength\tabcolsep{10pt}
  \begin{tabular}{ccc|ccc|ccc}
    \hline
    \hline    
    \multicolumn{3}{c|}{SM (a)} & \multicolumn{3}{c|}{SM (b)} & \multicolumn{3}{c}{SM (c)} \\
    $J^\pi$ & $E$ (keV) & $C^2S$  & $J^\pi$ & $E$ (keV)  & $C^2S$ & $J^\pi$ & $E$ (keV) & $C^2S$\\		
    \hline                        
    $0_1^+$ &    0 & 1.742  & $0_1^+$ &    0 & 1.575 & $0_1^+$ &    0 & 1.454 \\  
    $0_2^+$ & 2271 & 0.056  & $0_2^+$ & 1691 & 0.098 & $0_2^+$ &  444 & 0.105 \\  %
    $0_3^+$ & 3066 & 0.001  & $0_3^+$ & 2034 & 0.006 & $0_3^+$ & 1483 & 0.052 \\  %
    \hline
    $1_1^+$ & 2116 & 0.823  & $1_1^+$ & 1961 & 0.725 & $1_1^+$ & 2048 & 0.671 \\  
    \hline
    $2_1^+$ & 1959 & 0.829  & $2_1^+$ & 1662 & 0.402 & $2_1^+$ &  705 & 0.002 \\  
    $2_2^+$ & 2307 & 0.001  & $2_2^+$ & 1905 & 0.246 & $2_2^+$ & 1442 & 0.061 \\  %
    $2_3^+$ & 2706 & 0.064  & $2_3^+$ & 2155 & 0.035 & $2_3^+$ & 1804 & 0.013 \\  %
    $2_4^+$ & 2884 & 0.014  & $2_4^+$ & 2160 & 0.061 & $2_4^+$ & 1883 & 0.378 \\  %
    \hline
    $3_1^+$ & 2345 & 0.828  & $3_1^+$ & 2078 & 0.699 & $3_1^+$ & 1885 & 0.517 \\  
    \hline
    $4_1^+$ & 2250 & 0.134  & $4_1^+$ & 2011 & 0.038 & $4_1^+$ & 1326 & 0.002 \\  
    $4_2^+$ & 2278 & 0.811  & $4_2^+$ & 2120 & 0.720 & $4_2^+$ & 1818 & 0.541 \\  %
   \hline
   \hline
  \end{tabular} 
  \caption{Comparison of \dpsrf{95}{96} spectroscopic factors and excitation energies from the shell model calculations. The labels SM (a), (b) and (c) denote the three proton model spaces that were investigated (see text).}
  \label{tab:ShellModel96}
\end{center}
\end{table*}
In the \dpsrf{95}{96} reaction each state with $J>0$ can be populated by more than one value for the angular momentum transfer. The coupling of the $1/2^+$ ground state of \nuc{95}{Sr} to a valence neutron in $1d_{5/2}$ ($J=2,3$), $2s_{1/2}$ ($J=0,1$), $1d_{3/2}$ ($J=1,2$), and $0g_{7/2}$ ($J=3,4$) leads to various final states. The shell model calculations suggest that $1d_{5/2}$ dominates the $J=2,3$ states and the contribution of $2s_{1/2}$ to the $1^+$ states is negligible. Indeed the experimental angular distributions for the $1^+$ candidates are welled accounted for by $\Delta \ell =2$ transfer as shown in Fig.~\ref{fig:AngDist96_L2}. The results of the calculations are compared to the experimental data in Fig.~\ref{fig:ShellModel96}.

According to the calculations, the ground state of \nuc{96}{Sr} is dominated ($>60\%$) by a neutron $(1d_{5/2})^6(2s_{1/2})^2$ configuration with substantial ($\sim15\%$) $(1d_{5/2})^4(2s_{1/2})^2(1d_{3/2})^2$ contributions in all of the model spaces. The transfer from the $1/2^+$ ground state of \nuc{95}{Sr} has, therefore, a large spectroscopic factor approaching that of the independent particle model ($C^2S=2$). The result depends only weakly on the proton model space, reflecting the result obtained for \nuc{95}{Sr} where the \sfac\ of the $1/2^+$ ground state (and the $3/2^+$ first excited state) only weakly depend on the available proton space.
The predicted spectroscopic factor ($C^2S_\text{SM}\sim1.5$) was found to be much larger than the experimental result ($C^2S_\text{exp}=0.19(3)$), suggesting that the ground state of \nuc{96}{Sr} can not be well-described within the context of the spherical shell model.
Assuming axial symmetry a Coulomb excitation experiment determining the quadrupole moment of the $2^+_1$ state suggests a weakly deformed ($\beta\sim0.1$) ground state~\cite{Clement2016,Clement2016a}.

On the other hand, the experimental spectroscopic factors for the excited \zps\ are substantially larger than for the ground state. 
The 1229 and 1465~keV states in \nuc{96}{Sr} are known to arise from the mixing between a strongly deformed and a nearly spherical configuration, as evidenced by the large $\rho^2(E0)$ transition strength between them~\cite{Jung1980}. 
The strongly deformed states should not be populated directly in one-neutron transfer onto the spherical \nuc{95}{Sr} ground state. Therefore, the \sfacs\ of these states reflects their underlying spherical component which is populated strongly by the $(d,p)$ reaction. This suggests the existence of three different shapes in \nuc{96}{Sr}, with a weakly deformed, likely oblate, ground state and strongly mixed spherical and well deformed (prolate with $\beta =0.31(3)$) configurations in the excited $0^+$ states. This is discussed in more detail in Ref.~\cite{cruz18}. 

Given that the ground state of \nuc{96}{Sr} was not well reproduced in any of the calculations, it is expected that there will also be substantial discrepancies with the low energy states of \nuc{96}{Sr}. 
The wave function for the $2_1^+$ state was predicted to be dominated by the neutron $(1d_{5/2})^6(2s_{1/2})^1(1d_{3/2})^1$ configurations in shell model calculation (a) ($73\%$) and (b) ($27\%$), which has a large overlap with the \nuc{95}{Sr} ground state. Within the model space of calculation (c), many additional contributions were present in the lowest energy $2^+$ state and the spectroscopic factor (Table~\ref{tab:ShellModel96}) is very small. The drop in energy of the $2^+$ state to 705~keV in model (c) reflects the lowering of the $7/2^+$ state in \nuc{95}{Sr} as excitations to the proton $0g_{9/2}$ orbital become possible. 
The large spectroscopic factor predicted for the $2^+_4$ state reflects its wave function composition, which in this case is similar to the $2^+_1$ state of the other calculations. The experimental 2084~keV state might be associated with this level. In agreement with the experimental results, the calculations in model space (c) predict small spectroscopic factors for the other $2^+$ states. 
The first $2^+$ state in \nuc{90-96}{Sr} was previously interpreted as a proton spin-flip excitation from the $1p_{3/2}$ to the $1p_{1/2}$ orbital as no indications of the neutron sub-shell closure are visible at $N=56$.
The constant excitation energy can then explained by the quenching of the proton $1p_{3/2} - 1p_{1/2}$ spin-orbit splitting as the neutron $1d_{5/2}$ orbital is filled~\cite{FP1984}. Such configurations would not be populated here using the $(d,p)$ reaction. The small experimental spectroscopic factor for the $2^+$ state is consistent with a proton excitation or with a non-spherical configuration that has a small overlap with the \nuc{95}{Sr} ground state.


The main contributions to the wave function of the low-lying $4^+$ states are the neutron $(1d_{5/2})^5(2s_{1/2})^2(1d_{3/2})^1$ and $(1d_{5/2})^6(2s_{1/2})^1(0g_{7/2})^1$ configurations. The latter configuration can be populated directly via one-neutron transfer ($\Delta \ell=4$), which results in an enhancement of the spectroscopic factor as seen in Table~\ref{tab:ShellModel96}.
There is no strong evidence to suggest that the structure of the 1793~keV $4^+_1$ \nuc{96}{Sr} state is well-described within any of the present calculations. The $4^+$ state at 2120~keV has a larger spectroscopic factor, and may be associated with the calculated $4^+_1$ state. Additionally, $\Delta \ell = 4$ strength has been observed around $E=3200$~keV, but could not be assigned to a particular state~\cite{CruzThesis}.
A low-lying $3^+$ state was also predicted in each of the model spaces. The same $(1d_{5/2})^6(2s_{1/2})^1(0g_{7/2})^1$ configuration was found to be the primary component of this state, contributing 67\%, 47\% and 33\% to the total wave function in model spaces (a), (b), and (c), respectively. 
Experimentally, there is no candidate for a $3^+$ state with large spectroscopic factor, although the $4^+$ assignment of the 2120~keV state is tentative, and could be a $3^+$ state.
Another state of interest is the first $1^+$ state, which appears at around 2~MeV in all of the calculations. This state originates from the neutron $(1d_{5/2})^6(2s_{1/2})^1(1d_{3/2})^1$ configuration, which can be populated directly via $\Delta \ell=2$ transfer. The calculations predict that this configuration makes up 78\%, 68\% and 61\% of the total wave function in model spaces (a), (b), and (c), respectively. The $1^+$ state at 1995~keV is a likely candidate for this configuration, as it was strongly populated in the \dpsr{95}{Sr} reaction. 

To summarize, the spectroscopic strength in \nuc{96}{Sr} is smaller and more fragmented than in the shell model calculations, in particular for the $0^+$ and $2^+$ states. The absolute spectroscopic factors are not reproduced, but the rather large spectroscopic factors for low-lying $1^+$ and $4^+$ states are overall in line with the calculations. The discrepancy for the $0^+$ states, with the observation of the majority of the spectroscopic strength in the excited $0^+$ states, suggests that the ground state of \nuc{96}{Sr} is not spherical, but rather weakly (oblate) deformed~\cite{cruz18}.

\subsection{\nuc{97}{Sr}}
The comparison of the experimental results with the shell model calculations in Table~\ref{tab:ShellModel9597} suggest that the structure of \nuc{97}{Sr} is more complicated than for \nuc{95}{Sr}. The ground state spin and parity $1/2^+$~\cite{Buchinger1987} is unexpected in the framework of the spherical shell model, where the $2s_{1/2}$ orbital should be fully occupied at $N=59$. Isotope shift measurements across the Sr chain indicate that the ground state of \nuc{97}{Sr} is either spherical or weakly deformed~\cite{Buchinger1990}. The magnetic moment of the \nuc{97}{Sr} ground state is close to the value of \nuc{95}{Sr} and much smaller than the Schmidt value. The close-lying $0g_{7/2}$ and $1d_{3/2}$ $K^\pi = 1/2^+$ orbitals could lead to substantial mixing even for weakly deformed states, and thus explain these results.

In addition to the excitation energies, the calculated spectroscopic factors for the \dpsr{96} reaction are listed in Table~\ref{tab:ShellModel9597}.
As discussed previously, the striking discrepancies between the calculated spectroscopic factors for the \dpsr{95} reaction and our experimental results indicate that the shell model will not adequately describe the \dpsr{96} reaction. A good description of the \nuc{96}{Sr} ground state wave function is essential for calculating the overlap with states in \nuc{97}{Sr} and the results from the \dpsr{95} reaction make it clear that \nuc{94}{Sr} and \nuc{95}{Sr} ground states are well described by the shell model but the \nuc{96}{Sr} ground state is not. The interpretation of the spectroscopic factors is thus limited here to qualitative remarks.

From the weak population of the \nuc{97}{Sr} ground state in the \dpsr{96} reaction we can conclude that it has a considerably different wave function than that of the weakly deformed \nuc{96}{Sr} ground state, although this does not necessarily imply that it is strongly deformed. Clearly, further experimental measurements must be made to elucidate the structure of this state.
The largest spectroscopic factor is found here for the $3/2^+$ state, similar to \nuc{95}{Sr}, yet this state does not necessarily have the same structure as the configuration of the even-even projectile affects the spectroscopic factor as well. Relatively strong population of a low-lying $5/2^+$ state via the \dpsr{96} reaction indicates that there are substantial vacancies in the neutron $1d_{5/2}$ orbital in the \nuc{96}{Sr} ground state and this level could be regarded as the $N=59$ analogue of the 681~keV $5/2^+$ \nuc{95}{Sr} state.

\section{Summary and Outlook}
In summary, states in \nuc{95,96,97}{Sr} have been studied via the \dpsr{94,95,96} reactions for the first time. In total, 16 angular distribution measurements and associated spectroscopic factors have been determined. Spectroscopic factors were deduced for an additional 2 states by using a relative $\gamma$-ray analysis. These spectroscopic factors were compared to shell model calculations using realistic effective interactions within several carefully chosen valence spaces.

In \nuc{95}{Sr}, firm spin and parity assignments of $3/2^+$, $7/2^+$ and $5/2^+$  have been made for the 352, 556 and 681~keV states, respectively. Further constraints on the spin and parity of the 1666~keV state have been made, based on predicted cross sections. Good agreement was observed between experiment and shell model calculations, which suggests that low-lying states in \nuc{95}{Sr} arise from relatively simple neutron configurations.

In \nuc{96}{Sr}, all angular distribution analyses that were carried out confirm and refine previous spin and parity assignments, and new spin and parity constraints of $1^+,2^+,3^+$ have been made for the 2576 state.
A state at 3506(5)~keV has been newly identified, which is a candidate for a $1^+$ or $2^+$ level.
It was found that the excited \zps\ possess a larger overlap with the ground state of \nuc{95}{Sr} than the $0^+_1$ state, as evidenced by the larger spectroscopic factors. This result is in contrast to the shell model calculations, which predict that almost all of the $\Delta \ell=0$ strength is concentrated in the $0^+_1$ state.
A weakly deformed structure is suggested for the \nuc{96}{Sr} ground state. The results presented here also agree with the proposed proton configuration of the $2^+_1$ state~\cite{FP1984} which is not strongly populated in the present experiment. 

In \nuc{97}{Sr}, substantial spectroscopic strength to the 167 and 522~keV states was observed while the ground state was very weakly populated. The angular distributions are in agreement with the established spins and parities of the 167 and 522~keV states, however no quantitative comparison with the shell model could be made as the \nuc{96}{Sr} ground state was not well-described within the calculations.

The results discussed here provide valuable information concerning the single-particle composition of states in \nuc{95,96,97}{Sr}. By comparing the experimental spectroscopic factors to shell model calculations, we are able to gain an improved understanding of structural changes that indicate a departure from simple shell structure for $N\geq58$. In future, two-neutron transfer reactions should provide for a complementary examination of the underlying structure of the $0^+$ states in the even-even neutron-rich \nuc{}{Sr} isotopes. Low-energy Coulomb excitation to characterize the deformation of excited states in the even-odd Sr nuclei could provide information complementary to the present work. Lastly, large-scale shell model calculations in larger valence spaces, which have been so far only applied to the neutron-rich Zr isotopes~\cite{Sieja2009,Togashi2016}, will provide an important addition to the present discussion. 



\acknowledgments
The efforts of the TRIUMF operations team in supplying the \nuc{}{Sr} beams are highly appreciated. We acknowledge support from the Science and Technologies Facility Council (UK, grants EP/D060575/1 and ST/L005727/1), the National Science Foundation (US, grant PHY-1306297), the Natural Sciences and Engineering Research Council of Canada, the Canada Foundation for Innovation and the British Columbia Knowledge and Development Fund. TRIUMF receives funding via a contribution through the National Research Council Canada. N.A.O. acknowledges support from the CNRS PICS ``PACIFIC''.

\bibliography{references}

\end{document}